\def\comments{off}\def\arxiv{on}\def\showlines{off}
\ifdefined\comments\else\def\comments{off}\fi
\ifdefined\arxiv\else\def\arxiv{on}\fi
\ifdefined\showlines\else\def\showlines{on}\fi

\documentclass[a4paper,USenglish,cleveref,numberwithinsect,thm-restate]{lipics-v2021}

\makeatletter
\let\orig@GenericWarning\GenericWarning
\newif\iflocal@skip
\newcommand{\local@skipWarn}[4]{\edef\local@haystack{\detokenize{#1}}\edef\local@needle{\detokenize{#2}}\typeout{SKIP PACKAGE|\local@haystack|\local@needle|}\IfSubStr{\local@haystack}{\local@needle}{\edef\local@haystack{\detokenize{#3}}\edef\local@needle{\detokenize{#4}}\typeout{SKIP MESSAGE|\local@haystack|\local@needle|}\IfSubStr{\local@haystack}{\local@needle}{\local@skiptrue }{}}{}}

\def\GenericWarning#1#2{\local@skipfalse \local@skipWarn{#1}{class}{#2}{reference format}\local@skipWarn{#1}{class}{#2}{ACM keywords}\local@skipWarn{#1}{class}{#2}{CCS concepts}\local@skipWarn{#1}{fancyhdr}{#2}{headheight}\local@skipWarn{#1}{balance}{#2}{might not be balanced}\iflocal@skip\else \orig@GenericWarning{#1}{#2}\fi }
\makeatother

\usepackage[utf8]{inputenc}

\usepackage{bm}
\usepackage[shortlabels,inline]{enumitem}
\usepackage{mathtools}
\usepackage{subcaption}
\usepackage{tcolorbox}
\usepackage{thmtools}
\usepackage{tikz}
\usepackage{xfrac}

\newtcolorbox{todobox}{
  colback=yellow!20, colframe=red!80!black, boxrule=1pt, arc=4pt, left=4pt, right=4pt, top=2pt, bottom=2pt, fonttitle=\bfseries,
  title=TODO
}

\newcommand{\todo}[1]{\ifdim\lastskip>0pt\ignorespaces\fi}
\newcommand{\sharon}[1]{\ifdim\lastskip>0pt\ignorespaces\fi}
\newcommand{\neta}[1]{\ifdim\lastskip>0pt\ignorespaces\fi}
\newcommand{\proofcomment}[1]{\ifdim\lastskip>0pt\ignorespaces\fi}

\ifdefstring{\comments}{on}{
\RenewDocumentCommand{\sharon}{+m}{{\textcolor{purple}{\textbf{SS:} {\em #1}}}}
\RenewDocumentCommand{\neta}{+m}{{\textcolor{blue}{\textbf{NE:} {\em #1}}}}
\RenewDocumentCommand{\proofcomment}{+m}{{\textcolor{gray}{\em #1}}}
\RenewDocumentCommand{\todo}{+m}{\begin{todobox}#1\end{todobox}}
}{}

\newif\ifarxiv
\ifdefstring{\arxiv}{on}{\arxivtrue}{\arxivfalse}

\ifdefstring{\showlines}{on}{\linenumbers}{\nolinenumbers}

\def\removeFirst#1{\edef\temp{#1}\expandafter\removeFirstAux\temp\relax
}
\def\removeFirstAux#1#2\relax{#2}

\NewDocumentCommand{\NewAdjustableFunction}{mm}{\NewDocumentCommand{#1}{t'e{^_}d()}{#2\IfBooleanT{##1}{'}\IfNoValueF{##2}{^{##2}}\IfNoValueF{##3}{_{##3}}\IfNoValueF{##4}{\parens{\IfBlankTF{##4}{\cdot}{##4}}}}}

\NewDocumentCommand{\NewAdjustableOperator}{mm}{\NewAdjustableFunction{#1}{\operatorname{#2}}}

\NewDocumentCommand{\mathsc}{m}{{\normalfont\textsc{#1}}}

\DeclarePairedDelimiter{\parens}{\lparen}{\rparen}
\DeclarePairedDelimiter{\braces}{\lbrace}{\rbrace}
\DeclarePairedDelimiter{\angles}{\langle}{\rangle}
\DeclarePairedDelimiter{\bracks}{\lbrack}{\rbrack}

\DeclarePairedDelimiterX{\slashes}[1]
  {/}
  {/}
  {\mkern-1mu#1\mkern-1mu}

\DeclarePairedDelimiter{\abs}{\lvert}{\rvert}
\DeclarePairedDelimiterX{\midbraces}[2]{\lbrace}{\rbrace}{\,#1\nonscript\:\delimsize\vert \allowbreak \nonscript\:\mathopen{}#2\,}

\makeatletter
\NewCommandCopy{\old@mids}{\midbraces}
\RenewDocumentCommand{\midbraces}{somm}{\IfBooleanTF{#1}{\old@mids{#3}{#4}}{\IfNoValueTF{#2}{\old@mids*{#3}{#4}}{\old@mids[#2]{#3}{#4}}}}

\NewDocumentCommand{\swap@paired}{m}{\ExpandArgs{cc}\NewCommandCopy{old@#1}{#1}\ExpandArgs{c}\RenewDocumentCommand{#1}{som}{\IfBooleanTF{##1}{\UseName{old@#1}{##3}}{\IfNoValueTF{##2}{\UseName{old@#1}*{##3}}{\UseName{old@#1}[##2]{##3}}}}}
\forcsvlist{\swap@paired}{parens,braces,angles,bracks,bbracks,slashes,abs}
\makeatother

\newcommand{\theory}{\mathcal{T}}

\NewDocumentCommand{\NewTheoryCommand}{mm}{\NewDocumentCommand{#1}{se{^_}d()}{#2\IfBooleanT{##1}{^{\theory}}\IfNoValueF{##2}{^{##2}}\IfNoValueF{##3}{_{##3}}\IfNoValueF{##4}{\parens{\IfBlankTF{##4}{\cdot}{##4}}}}}

\NewTheoryCommand{\sig}{\Sigma}
\NewTheoryCommand{\Sorts}{\mathcal{S}}
\NewTheoryCommand{\Terms}{\mathsc{Terms}}
\NewTheoryCommand{\qf}{\mathsc{QF}}
\NewTheoryCommand{\lang}{\mathcal{L}}

\makeatletter
\edef\orig@models{\models}
\RenewDocumentCommand{\models}{so}{\IfBooleanTF{#1}{\models[\theory]}{\IfNoValueTF{#2}{\orig@models}{\models^{#2}}}}
\makeatother

\newcommand{\llt}{\ensuremath{\prec}}
\newcommand{\lgt}{\succ}

\newcommand{\liff}{\leftrightarrow}

\NewAdjustableOperator{\ite}{ITE}

\NewTheoryCommand{\domain}{\mathcal{D}}
\NewTheoryCommand{\interp}{\mathcal{I}}

\newcommand{\Land}{\bigwedge}

\NewDocumentCommand{\sort}{}{\bm{s}}
\NewDocumentCommand{\sinf}{}{\sort^\infty}

\NewDocumentCommand{\NewMathAcronym}{mm}{\NewDocumentCommand{#1}{}{\ifmmode\mathrm{#2}\else #2\fi
  }}

\newcommand{\frag}{\mathcal{F}}

\NewMathAcronym{\epr}{EPR}
\NewMathAcronym{\fol}{FOL}
\NewMathAcronym{\fsf}{SF}
\NewMathAcronym{\wsis}{WS1S}

\NewDocumentCommand{\Description}{+m}{}
\usetikzlibrary{arrows}
\usetikzlibrary{calc}
\usetikzlibrary{decorations.markings}
\usetikzlibrary{fit}
\usetikzlibrary{positioning}
\newlength{\summaryDouble}
\newlength{\nodeSize}
\tikzset{
  symbolic/.style = {
    x=1.25cm, 
    y=1cm, 
    node distance=1.5cm,
    font=\small,
  },
  element/.style = {
    circle, draw,
    minimum size=\nodeSize,
    inner sep=0pt,
    font=\small,
  },
  regular/.style = {element, solid},
  explicit/.style = {element, dotted},
  summary/.style = {
    element, 
    double, 
    double distance=\summaryDouble,
    rectangle, rounded corners,
  },
  summaryFit/.style = {
    draw, 
    rounded corners, 
    double, 
    double distance=\summaryDouble,
    inner ysep=5pt,
  },
  dummy/.style = {
    element,
    draw=none
  },
  arrow/.style = {
    draw,
    ->,
    >=stealth',
    semithick,
    shorten >=1pt,
    shorten <=1pt,
  },
  darrow solid/.style={
    arrow
  },
  darrow dots/.style={
    draw,
    dash pattern=on 0pt off 6pt,
    line cap=round,
    line width=2pt
  }
}

\tikzset{
	tridots/.pic={
		\begin{scope}[
			x=0.25cm, y=0.25cm, 
			shift={(0,-3.5)}, 
			scale=1, 
			transform shape
		]
		\node at (0,0) {\Huge $\cdot$};
		\node at (-1,0) {\Huge $\cdot$};
		\node at (1,0) {\Huge $\cdot$};

		\node at (0,1.732) {\Huge $\cdot$};

		\node at (-0.5, 0.866) {\Huge $\cdot$};
		\node at (0.5, 0.866) {\Huge $\cdot$};
		\end{scope}
	}
}
\newcommand{\tridots}{\begin{tikzpicture}[baseline={(0,-3.25ex)}]
	\pic[scale=0.55] at (0,0) {tridots};
\end{tikzpicture}}

\NewDocumentCommand{\scalepic}{+m}{\scalebox{0.85}{#1}}

\ExplSyntaxOn
\NewDocumentCommand{\sub}{ r[] }{
  \bracks*{
    \seq_set_split:Nnn \l_tmpa_seq { , } { #1 }
    \seq_map_indexed_inline:Nn \l_tmpa_seq { 
      \__sub_process_item:n { ##2 }
      \int_compare:nNnF { ##1 } = { \seq_count:N \l_tmpa_seq }
        { , }
    }
  }
}
\cs_new_protected:Nn \__sub_process_item:n {
  \tl_if_blank:nTF { #1 } { 
    \dots 
  }{
    \__sub_split_pair:n #1 \q_stop
  }
}
\cs_new_protected:Nn \__sub_split_pair:n {
  \__sub_split_pair_aux:w #1
}
\cs_new_protected:Npn \__sub_split_pair_aux:w #1/#2\q_stop {
  #2 \mapsto #1
}
\ExplSyntaxOff

\NewDocumentCommand{\bracksIf}{mO{\llt}}{\IfNoValueTF{#1}{}{\bracks{\IfBlankTF{#1}{#2}{#1}}}}
\NewDocumentCommand{\strict}{}{\mathsc{strict}}
\NewDocumentCommand{\total}{}{\mathsc{tot}}
\NewDocumentCommand{\pref}{}{\mathsc{pref}}
\NewDocumentCommand{\psucc}{}{\mathsc{prosucc}}
\NewDocumentCommand{\rpred}{}{\mathsc{regpred}}

\NewDocumentCommand{\ordax}{}{\alpha}
\NewDocumentCommand{\osc}{so}{\ifmmode \mathrm{OSC}\else OSC\fi \ensuremath{\IfBooleanT{#1}{^{\star}}\bracksIf{#2}[\ordax]}}

\NewTheoryCommand{\M}{\mathcal{M}}

\NewAdjustableOperator{\Th}{Th}

\NewDocumentCommand{\lia}{}{{\ifmmode \mathchoice{\mathrm{LIA}}{\mathrm{LIA}}{\mathsc{lia}}{\mathsc{lia}}\else LIA\fi }}
\newcommand{\intSort}{\texttt{\bfseries int}}

\NewDocumentCommand{\str}{}{{\ifmmode \mathchoice{\mathrm{STR}}{\mathrm{STR}}{\mathsc{str}}{\mathsc{str}}\else STR\fi }}

\newcommand{\Sreg}{\ensuremath{\bm{S}_\mathrm{reg}}}
\NewDocumentCommand{\Reg}{sO{\Delta}}{\IfBooleanTF{#1}{\mathcal{R}}{\Reg*_{#2}}}

\NewDocumentCommand{\regexp}{m}{\slashes{#1}}
\newcommand{\stringSort}{\texttt{\bfseries string}}

\NewAdjustableOperator{\prefOf}{pref}
\NewAdjustableOperator{\trim}{trim}
\newcommand{\preflt}{\sqsubset}

\NewDocumentCommand{\R}{t*t+t.t-}{\IfBooleanT{#1}{$^*$}\IfBooleanT{#2}{$^+$}\IfBooleanT{#3}{$\cdot\mkern 2mu$}\IfBooleanT{#4}{|$\cdots\mkern-0.1mu$|}}

\NewDocumentCommand{\nregexp}{m}{\begingroup \slashes[\big]{\texttt{#1}}\endgroup }

\NewAdjustableFunction{\bound}{\mathcal{B}}
\NewAdjustableFunction{\explic}{\mathcal{E}}
\NewDocumentCommand{\expel}{r<>}{\angles{#1}}

\newcommand{\treg}{t^{\operatorname{reg}}}
\newcommand{\AllAtoms}{\mathbb{A}}
\newcommand{\mixed}{\mathbb{T}}
\NewAdjustableFunction{\elcls}{\tau}
\newcommand{\atom}{\gamma}
\NewAdjustableFunction{\atoms}{\Gamma}

\NewAdjustableFunction{\segS}{C_\llt}
\NewAdjustableFunction{\segL}{C_\lgt}
\NewAdjustableOperator{\seg}{seg}

\NewAdjustableOperator{\treePos}{pos}
\NewAdjustableOperator{\treeNegSpine}{neg-sp}
\NewAdjustableOperator{\treeNegBranch}{neg-br}
\NewAdjustableOperator{\treeNegRoot}{neg-rt}
\NewAdjustableOperator{\treeSameRoot}{eq-rt}

\NewAdjustableOperator{\treeParent}{parent}
\NewAdjustableOperator{\treeChild}{child}
\NewAdjustableOperator{\treeIsChild}{is-child}
\NewAdjustableOperator{\treeSibling}{sibling}

\newcommand{\tif}{\text{if }}
\newcommand{\tand}{\text{ and }}
\newcommand{\tother}{\text{otherwise}}

\AtEndPreamble{
  \theoremstyle{remark}

  \theoremstyle{plain}

  \theoremstyle{definition}
  \newtheorem{desired}[theorem]{Desired Property}
  \crefname{desired}{Desired Property}{Desired Properties}

  \newlist{citemize}{itemize}{1}
  \setlist[citemize]{label=\Large\textbullet,left=7pt}
  \newlist{cdescription}{description}{1}
  \setlist[cdescription]{labelindent=7pt,
    leftmargin=14pt
  }
}

\newcommand{\fest}{\texttt{FEST}}
\newcommand{\mona}{\texttt{MONA}}

\newcommand{\defeq}{\triangleq}

\newcommand{\Int}{\mathbb{Z}}
\newcommand{\Mid}{\;\big\vert\;}

\NewDocumentCommand{\para}{O{.}m}{\smallbreak \noindent\textbf{#2#1}\enskip }

\NewDocumentCommand{\code}{m}{\text{\lstinline|#1|}}

\newcommand{\Proofs}{}

\NewDocumentCommand{\ProofOf}{m+m}{\begin{proof}[Proof of \Cref{#1}]\phantomsection
		\label{proof:#1}
		#2
	\end{proof}}

\NewDocumentCommand{\Proof}{m+m}{\proofcomment{proof deferred to \Cref{sec:proofs}: \hyperref[proof:#1]{proof}}\toks0=\expandafter{\Proofs}\toks2=\expandafter{\ProofOf{#1}{#2}}\xdef\Proofs{\the\toks0 \the\toks2}}

\makeatletter
\newcommand{\namedlabel}[2]{\phantomsection \def\@currentlabelname{#2}\label{#1}}
\makeatother \title{Decidability Results for Fragments 
  of First-Order Logic
  via a Symbolic Model Property}
\titlerunning{Decidability Results for Fragments 
  of FOL
  via a Symbolic Model Property}
\relatedversiondetails[cite=elad26-decidability-arxiv]{Extended Version}{https://doi.org/10.48550/arXiv.2605.05840}

\author{Neta Elad}{Tel Aviv University, Israel}{netaelad@mail.tau.ac.il}{https://orcid.org/0000-0002-5503-5791}{}

\author{Sharon Shoham}{Tel Aviv University, Israel}{sharon.shoham@gmail.com}{https://orcid.org/0000-0002-7226-3526}{}

\Copyright{Neta Elad and Sharon Shoham}
\authorrunning{Neta Elad and Sharon Shoham}

\ccsdesc[500]{Theory of computation~Logic and verification}
\keywords{first-order logic, decidability, symbolic structures}

\EventEditors{Claudia Faggian and Joost-Pieter Katoen}
\EventNoEds{2}
\EventLongTitle{41st Annual Symposium on Logic in Computer Science (LICS 2026)}
\EventShortTitle{LICS 2026}
\EventAcronym{LICS}
\EventYear{2026}
\EventDate{July 20--23, 2026}
\EventLocation{Lisbon, Portugal}
\EventLogo{}
\SeriesVolume{380}
\ArticleNo{35}

\acknowledgements{The research leading to these results has received funding from the European Research Council under the European Union's Horizon 2020 research and innovation programme (grant agreement No [759102-SVIS]). This research was partially supported by Israel Science Foundation (ISF) grant No.\ 2117/23.
} 
\begin{document}

\maketitle

\begin{abstract}\neta{comments check: arxiv is \arxiv}Recently, symbolic structures were proposed
as finite representations of potentially infinite first-order structures,
where Linear Integer Arithmetic terms and formulas
define the domain and interpretations of a structure.
We generalize symbolic structures to use
any base theory that admits a standard model.
Symbolic structures induce a symbolic model property,
which holds for a fragment of first-order logic
if every satisfiable formula in the fragment has a symbolic model.
The symbolic model property implies decidability,
since the model-checking problem for symbolic structures is decidable.
We use the symbolic model property to prove decidability
for several fragments that extend
the fragment of stratified formulas,
relaxing the quantifier-alternation constraints
by allowing one sort to have self-looping functions,
under certain restrictions.
To establish the symbolic model property for these fragments
we construct a symbolic model 
for a formula
from an arbitrary model.
The construction and its correctness are proved 
in a generic fashion,
which may be instantiated to other 
similarly restricted fragments. \end{abstract}

\section{Introduction}
In this paper we study a formalism for representing infinite models of formulas in first-order logic (\fol) and use this formalism to prove decidability results
for new fragments of \fol.
Our investigation is motivated by the use of \fol{}
for verification.
First-order logic offers an expressive language
for specifying verification conditions,
logical formulas that capture the correctness
of a computer program or system,
as well as means for automation,
due to its complete proof system.

In particular, we focus on the use of
quantified \fol{} formulas
for modeling systems with unbounded domains,
where axiomatically-defined order relations
are used to abstract important primitives that are not directly definable in \fol,
e.g., reachability in linked lists
or tree-shaped data-structures.
\fol\ offers a natural way
to express verification conditions of such systems,
allowing the user to write specifications
that match their mental model of the system,
and verify the correctness of the system
(see, e.g.,~\cite{linked-lists-epr,implicit-rankings,fo-functional-programs,modularity-epr-verification}).

However, \fol\ is undecidable in general,
and thus the verification process may fail.
Completeness of \fol\ does guarantee
that valid formulas admit finite proofs,
but there are no such guarantees on counter-models
of invalid formulas.
An \fol-based verification tool
may get stuck when trying to verify a system
due to counter-models it is unable to find,
leaving the user with no actionable feedback.
This is particularly common during intermediate steps
in the verification process,
when the user still refines the system's specifications
and fixes bugs in the system.

One approach to handle these issues is by limiting
verification conditions to a fragment of \fol\ that enjoys a finite model property, which means that every satisfiable formula admits a finite model.
The finite-model property implies decidability,
and, importantly, ensures that whenever a counter-model exists,
it may be found and presented to the user,
providing much-desired feedback.

One very successful fragment in this regard is
the fragment of stratified formulas
(\fsf)~\cite{paxos-made-epr},
the many-sorted variant of the
Effectively PRopositional fragment
(\epr)~\cite{bsr-epr,epr-decidable}.
The finite model property of \fsf\ is guaranteed by
requiring acyclicity of the
quantifier-alternation graph
of each formula,
where vertices correspond to sorts
and edges correspond to
$\forall\exists$ quantifier alternations
(either explicit or implicit via function symbols).
Though useful~\cite{paxos-made-epr,ivy,stellar-epr},
the fragment of \fsf\
is rather limiting in its syntactic constraints,
and may require nontrivial amount of work from the user
to fit specifications of complex systems
into the fragment~\cite{cex-driven-quant-inst}.

In this paper,
we pursue an alternative approach
that goes beyond a finite model property.
We explore effective representations of
potentially infinite models,
and use them to establish decidability
results
for fragments of \fol{}
that do not admit a finite model property.

We build on the notion of \emph{symbolic structures},
introduced in~\cite{infinite-needle} as finite representations
of potentially infinite models.
Symbolic structures use
the standard model of the integers
and the theory of Linear Integer Arithmetic (\lia)
to represent first-order structures.
In symbolic structures,
the domain is given symbolically
by finitely many nodes,
each of which represents
a potentially infinite set of integers,
and interpretations of signature symbols
are defined with terms and formulas in the theory of \lia.
This is akin to first-order interpretations
(over the standard model of the integers)
with similarly-behaved elements explicitly grouped together
into symbolic nodes.

We start by generalizing
the formalism of symbolic structures
to other theories,
producing a definition that is parametric by the
\emph{base theory} underlying the symbolic structure.
Our generalization allows
any base theory that admits a standard model
(i.e., any theory of a structure).
Given such a base theory,
the nodes in our generalized symbolic structures
represent sets of elements of the standard model of the theory,
and interpretations are defined using terms and
formulas in the language of the theory.
This generalization matches how first-order interpretations
may be defined over any base structure.

A key property of symbolic structures is that they allow
reducing model-checking,
which determines whether a given symbolic structure satisfies
a given first-order formula,
to validity of some formula in the base theory.
Thus, model-checking is decidable for symbolic structures
over decidable base theories
(which is unsurprising due to the connection between symbolic structures
and first-order interpretations).
Decidability of model-checking
induces a procedure
for providing counter-models to verification conditions
by enumerating and checking symbolic structures.
This procedure may not always halt,
since undecidability of validity in \fol{}
implies that some invalid formulas
do not have symbolic counter-models.

Our main contribution is proving decidability
for new fragments of \fol\ via
a \emph{symbolic} model property,
which states that every satisfiable formula has a symbolic model.
Due to decidability of model-checking for symbolic structures,
the symbolic model property trivially implies decidability.
However, establishing the symbolic model property may be difficult
and is the major problem this paper tackles.
Interestingly, and in contrast to
the fragment of stratified formulas,
the fragments we consider do not enjoy a finite model property,
and some formulas in them may only admit infinite models.
Still, the symbolic model property ensures that when verification fails,
a symbolic counter-model can be produced.

To define the decidable fragments,
we introduce the
Ordered Self-Cycle (\osc) family of fragments,
where each fragment, $\osc[]$,
is parameterized by an axiomatization
$\ordax$
of an order relation $\llt$
over some sort $\sinf$.
$\osc[]$
extends the fragment of stratified formulas
by relaxing the acyclicity requirement
of the quantifier-alternation graph,
allowing function symbols that introduce self-cycles over sort $\sinf$ in the graph,
under certain restrictions.
The $\osc$ family is a generalization
of the decidable fragment presented
in~\cite{infinite-needle},
which corresponds to $\osc[\total]$,
where $\total$ axiomatizes a total order.
Decidability of $\osc[\total]$
is proved in~\cite{infinite-needle}
via a symbolic model property,
using \lia{} as the base theory.

In this paper,
we establish a symbolic model property
for additional fragments
in the \osc{} family,
using both \lia{} and a decidable theory of strings
as base theories.
Most notably,
we use the base theory of strings
to show the symbolic model property of
the Prefix-Ordered Self-Cycle fragment,
$\osc[\pref]$,
where $\pref$ axiomatizes a prefix order,
a common construct in verification
that arises naturally when modeling reachability
in acyclic tree-shaped data-structures
or in acyclic linked lists with sharing.
We further prove symbolic model properties
for several variants of
$\osc[\total]$ and $\osc[\pref]$.

Our proofs
establish a symbolic model property
through a generic proof recipe,
which extends and simplifies
the proof given in~\cite{infinite-needle}
for $\osc[\total]$.
The proofs work by constructing
a symbolic model for a formula
from an arbitrary model of the formula.
The construction is defined in several steps,
some of which are general,
while others depend on the specific
order axiom $\ordax$ and base theory used.

Finally, we note that we have implemented
a proof-of-concept tool
for exploring generic symbolic structures
by extending the open-source
\fest{}~\cite{infinite-needle} Python library.
Our implementation generalizes \fest{}
to support different base theories
in a modular way,
in particular enabling model-checking
of symbolic structures over the base theory of strings.

In summary, this paper offers the following contributions:
\begin{enumerate*}[(1)]
  \item a general definition of symbolic structures,
    finite representations of potentially
    infinite models that can be viewed and acted
    upon~(\Cref{sec:symstructs});
  \item the introduction of the family of
    Ordered Self-Cycle
    fragments,
    which are parameterized by an
    order axiomatization~(\Cref{sec:osc}); and
\item decidability proofs
    for several \osc{} fragments,
    in particular the
    Prefix-Ordered Self-Cycle fragment
    $\osc[\pref]$,
    as well as variants of $\osc[\pref]$
    and $\osc[\total]$~(\Cref{sec:dec}).
\end{enumerate*}
The rest of the paper is organized as follows:
related work is discussed in~\Cref{sec:related},
preliminaries are given in~\Cref{sec:prelim},
\Cref{sec:impl} gives a brief overview of
the implementation of generic symbolic structures,
and \Cref{sec:conclusion} concludes.
\ifarxiv
We omit proofs of lesser salience 
and defer them to \Cref{sec:proofs}.
\else
Due to space considerations,
proofs of lesser salience are omitted.
See~\cite{elad26-decidability-arxiv}
for an extended version with detailed proofs.
\fi
 \section{Related Work}
\label{sec:related}
\para{Finite model property}
A long line of work has identified syntactic fragments of
\fol\
for which satisfiability is shown to be decidable
by establishing a finite model property.
Classical results include the
Ackermann, G{\"o}del, L{\"o}b
and Bernays-Sch{\"o}nfinkel-Ramsey fragments,
as well as later refinements and extensions
studied extensively in logic and finite model theory~\cite{ackermann-decidable,goedel-decidable,loeb1967-decidability,gurevich1976-decidable,mortimer1975-decidable,gurevich-book,decidable-many-sorted,kalmar-33,voigt-thesis}.
Though some of these fragments admit specialized
decision procedures,
they all enjoy a finite model property.
For a detailed survey, see~\cite{voigt-thesis}.
Related to this line of work
are decidability results for
satisfiability over finite structures
(e.g.,~\cite{shelah1977decidability,danielski-unfo,guarded-negation}).
All of these works are similar in spirit
to our symbolic-model-property approach,
but distinct as the fragments we consider
do not admit a finite model property.

\para{Decidability and infinite models}
The seminal works on
monadic second-order logic (MSO)~\cite{buchi-decidable-mso,buchi-mso-s1s,rabin-mso-s2s}
prove decidability for fragments
that do not admit a finite model property.
There, the specific structures of lines
(S1S) and trees (S2S) are considered,
and finite-state automata are used to represent
infinite models.
A complementary line of work studies
the construction of infinite models
via proof-theoretic methods~\cite{bachmair-original,infinite-models-clause-sets},
using saturated sets of clauses
to witness the existence of models.
Using the symbolic model property to prove decidability
of \fol{} fragments was introduced in~\cite{infinite-needle},
which this work generalizes
to other fragments of \fol{} and symbolic structures
over different base theories.

\para{Representing infinite structures}
Symbolic structures can also be expressed as
first-order interpretations (see, e.g.,~\cite{gradel15-fo-interp})
over the standard structure of the base theory,
by indirectly encoding the nodes in the symbolic domain
using the dimensionality of the FO interpretation.
The different nodes in a symbolic structure
provide a natural way to group and define
similarly-behaved elements.
An alternative approach to representing infinite structures
is by using automatic structures~\cite{blumensath04-finite,automatic-structures,automata-infinite-structures,gradel-finite-infinite},
where automata and regular languages are used.
However, these works do not provide a method to
automatically find a satisfying structure for a given formula,
nor identify a decidable fragment
where representable satisfying structures
are guaranteed to exist.
 \section{Preliminaries}
\label{sec:prelim}
\para{First-order logic}
We consider many-sorted first-order logic
with equality.
We use the usual definitions of
a signature $\sig$,
terms over signature $\Terms_\sig$,
formulas $\lang_\sig$,
first-order structures
and the satisfaction relation.
For a set of variables $X$,
we denote by
$\Terms_\sig(X)$ and
$\lang_\sig(X)$
the sets of terms and, respectively,
formulas over $\sig$
with free variables in $X$.
In particular,
we write $\Terms_\sig(\emptyset)$
for the set of ground terms
and $\lang_\sig(\emptyset)$ for
the set of
sentences (formulas without free variables)
over $\sig$.

\para{Theories of structures}
In this paper we consider
first-order theories of structures.
Given a first-order structure $\M$
for signature $\sig$,
the theory of $\M$,
denoted $\Th(\M)$,
is the set of all sentences
satisfied by $\M$,
i.e.,
$
\Th(\M) = \midbraces{
  \varphi \in \lang_\sig(\emptyset)
 }{ \M \models \varphi}
$.
For $\theory = \Th(\M)$,
we say that a sentence
$\varphi \in \lang_\sig(\emptyset)$
is \emph{valid} in $\theory$
(or $\theory$-valid),
denoted
$\models* \varphi$,
if $\varphi \in \theory$,
i.e.,
$\M \models \varphi$.
We say that a formula is
\emph{satisfiable} in $\theory$
(or $\theory$-satisfiable)
if its existential closure is $\theory$-valid.
We say that $\theory$
is \emph{decidable}
when there exists
an algorithm that decides
whether a sentence
is $\theory$-valid.

\para{The theory of Linear Integer Arithmetic (\lia)}
The theory of \lia\ is the theory
of the structure $\M^\lia$
for the signature
\[
\sig^\lia =
  \midbraces{ c_z }{ z \in \Int }
  \cup\braces{
    {+}(\cdot,\cdot)}
  \cup\braces{
    {<}(\cdot,\cdot)
  }
\]
over the single sort $\intSort$.
The domain of $\M^\lia$
consists of the integers
$\Int$,
and the interpretations of
the constant symbols $c_z$,
function symbol $+$
and relation symbol $<$
are defined in the usual way.
With abuse of notation,
we sometimes write $z$ for $c_z$.
Recall that the theory of \lia\
is decidable (see, e.g.,~\cite{lia-decidable}).
\todo{if we can, avoid abuse of notation}

\para{A decidable theory of strings}
Given a finite alphabet $\Delta$,
the language of regular expressions over $\Delta$,
denoted $\Reg$,
is defined by the following grammar:
\[
	r \in \Reg \Coloneq
		\epsilon
		\Mid \delta \in \Delta
		\Mid \parens{r_1 \cdot r_2}
		\Mid \parens{r_1 \mid r_2}
		\Mid \parens{r^*}.
\]
We further denote by
$\parens{r?}$
the regular expression
$\parens{r\mid \epsilon}$
and by
$\parens{r^+}$ the regular expression
$\parens{r \cdot \parens{r^*}}$.
We consider the decidable theory of strings
over $\Delta$,
denoted $\str_\Delta$,
which is
the theory of the structure
$\M^\str_\Delta$
(also known as
$\Sreg$~\cite{s-reg-left})
for the signature
\[
\sig^\str_\Delta =
	\braces{ c_\epsilon }
	\cup
	\midbraces{\ell_\delta(\cdot)}{\delta \in \Delta}
	\cup
	\braces{
		{\preflt}(\cdot,\cdot)
	}\cup
	\midbraces{
		\regexp{ r }(\cdot)
	}{r \in \Reg }
\]
over the single sort $\stringSort$.
The domain of $\M^\str_\Delta$
is the set of finite strings, also known as \emph{words},
over $\Delta$,
the constant symbol
$c_\epsilon$
is interpreted as the empty string
$\epsilon$,
the unary function symbols
$\ell_\delta$
(``last'')
are interpreted as
appending the letter $\delta$ to a string,
the binary relation symbol
${\preflt}$
is interpreted as the strict prefix relation over strings,
and the unary relation symbols
$\regexp{r}$
are
interpreted as
the languages (i.e., sets of words)
of the regular expressions $r$.
With abuse of notation we sometimes
use $\epsilon$ as a constant symbol
and $t \cdot \delta$ (for $\delta \in \Delta$) as
shorthand for $\ell_\delta(t)$.
As shown in~\cite{s-reg-left},
the theory $\str_\Delta$ is decidable
for any finite $\Delta$.
 \section{Symbolic Structures}
\label{sec:symstructs}
In this section we introduce a generalization
of the symbolic structures formalism~\cite{infinite-needle}.
Throughout the section we fix a theory
$\theory = \Th(\M*)$
of a structure
$\M* = \parens{ \domain*, \interp* }$
for a single-sorted
signature $\sig*$,
and define symbolic structures
over $\theory$.
We use the shorthand
$\lang*$ and $\Terms*$
for $\lang_{\sig*}$ and $\Terms_{\sig*}$
respectively.

Given a signature $\sig$ (not to be confused with $\sig*$), symbolic structures for $\sig$  are finite
representations of (explicit) first-order structures for $\sig$,
where the domain of each sort is stitched up
piecewise from subsets
of the domain of the standard model
$\M*$
of the underlying base theory
$\theory$; each such subset is summarized by a node in the symbolic structure.
Interpretations of constant, function and relation symbols from $\sig$
are given symbolically in symbolic structures,
using terms and formulas
in the language of $\sig*$.

The symbolic representation allows to check if the structure represented by a symbolic structure satisfies a formula $\varphi$ over $\sig$
by essentially substituting symbols in $\sig$ appearing in $\varphi$ by their symbolic definitions, transforming $\varphi$ into a formula over $\sig*$ that is $\theory$-valid iff the explicit structure satisfies $\varphi$.
Formally, symbolic structures
and their explications are defined
as follows.

\begin{definition}[Symbolic structure]
  A
  \emph{symbolic structure over
    base theory $\theory$}
  for a first-order signature $\sig$
  with sorts $\Sorts$
  is a triple
  $S = \parens{\domain^S, \bound^S, \interp^S}$,
  defined as follows.
  \begin{citemize}
    \item
    $\domain^S$ is
    a function from $\Sorts$
    to finite, non-empty sets
    of \emph{nodes},
    where
    $\domain^S(\sort_1) \cap \domain^S(\sort_2)
    = \emptyset$
    for every
    $\sort_1 \neq \sort_2 \in \Sorts$.
    We refer to
    $\bigcup_{\sort \in \Sorts} \domain^S(\sort)$
    as the \emph{symbolic domain} of $S$.
\item
    $\bound^S$
    is a function
    from the symbolic domain
    to $\theory$-satisfiable formulas
    in $\lang*(\braces{x})$
    (i.e.,
    formulas over $\sig*$ with at most
    one free variable $x$).
    For each node $n$ in the symbolic domain,
    $\bound^S(n)$ is called the
    \emph{bound formula} of $n$.
\item $\interp^S$
      is a function
      that maps
      each symbol $s \in \sig$
      to its \emph{symbolic interpretation}
      $\interp^S(s)$,
      also denoted as $s^S$.

      The symbolic interpretation of a
      constant symbol
      $c \colon \sort$
      is a pair
      $c^S = \expel<n,t>
      \in \domain^S(\sort) \times \Terms*(\emptyset)$
      such that
      \begin{equation}
        \label{eq:well-defined-c}
        \models* \bound(n) \sub[t/x].
      \end{equation}

      The symbolic interpretation
      of a function symbol
      $f \colon \sort_1 \times \dots \times \sort_m \to \sort$
      is a function
      $f^S
      \colon
      \domain^S(\sort_1) \times
      \dots \times \domain^S(\sort_m)
      \to
      \domain^S(\sort) \times
      \Terms*({
        \braces{x_1, \dots, x_m}
      })$
      such that
      whenever
      $f^S(n_1, \dots, n_m)
      = \expel<n,t>$,
      \begin{equation}
        \label{eq:well-defined-f}
        \models*
        \forall
        x_1, \dots, x_m.
        \parens{
          \Land_{j=1}^m \bound(n_j) \sub[x_j/x]
        } \to \bound(n) \sub[t/x].
      \end{equation}

      The symbolic interpretation of
      a relation symbol
      $R \colon \sort_1 \times \dots \times \sort_m$
      is a function
      $R^S
      \colon
      \domain^S(\sort_1) \times
      \dots \times \domain^S(\sort_m)
      \to
      \lang*({
        \braces{x_1, \dots, x_m}
      })$.
  \end{citemize}
  We call conditions
  \labelcref{eq:well-defined-c,eq:well-defined-f}
  the \emph{well-definedness}
  conditions of $S$.
\end{definition}

A symbolic structure $S$ is a finite representation
of an \emph{explicit} structure $\explic(S)$.
Next we define the explicit structure
represented by a symbolic structure.

\begin{definition}[Explication]
  Let
  $S = \parens{\domain^S, \bound^S, \interp^S}$
  be a symbolic structure
  over base theory $\theory$
  for a first-order signature
  $\sig$ with sorts $\Sorts$.
  The explication of $S$,
  denoted $\explic(S)$ is
  the first-order structure
  $\explic(S) =
  \parens{
    \domain^{\explic(S)},
    \interp^{\explic(S)}
  }$
  for $\sig$,
  defined as follows.
  \begin{citemize}
    \item
    The explicit domains $\domain^{\explic(S)}$ of sorts are given
    by explicating each node in the domain of a sort
    into a set of explicit elements from the domain of $\M*$ and taking the disjoint union of these sets.
    For a symbolic node $n$ we denote
    its explicit elements by
    \[
    \explic(n) \defeq
    \midbraces{ \expel<n,d> }
    {
      d \in \domain* \tand
        \M*, \sub[d/x] \models \bound(n)
    },
    \]
    and we define the explicit domain as
    \[
    \domain^{\explic(S)}(\sort) \defeq
    \bigcup_{n \in \domain^S(\sort)} \explic(n)
    \]
    \item
    The explicit interpretations $\interp^{\explic(S)}$ of symbols
    are given by explicating the symbolic interpretations
    according to $\M*$.

    The interpretation of a constant symbol
    $c \colon \sort \in \sig$
    where $c^S = \expel<n,t>$
    is the explicit element
    \[
    c^{\explic(S)} \defeq
    \expel<n, t^{\M*}>,
    \]
    where $t^{\M*}$
    is defined recursively
in the usual way.

    The interpretation of a function symbol
    $f \colon \sort_1 \times \dots \times \sort_m
    \to \sort \in \sig$
    is the function
    $
    f^{\explic(S)} \colon
    \domain^{\explic(S)}(\sort_1) \times \dots \times
    \domain^{\explic(S)}(\sort_m) \to
    \domain^{\explic(S)}(\sort)
    $, defined such that if
    $f^S(n_1, \dots, n_m) = \expel<n,t>$ then
    \[
    f^{\explic(S)}
    \parens{
      \expel<n_1, d_1>, \dots,
      \expel<n_m, d_m>
    } \defeq
    \expel<
      n,
      t^{\M*}_{\sub[d_1/x_1,,d_m/x_m]}
>.
    \]

    The interpretation of a relation symbol
    $R \colon \sort_1 \times \dots \times \sort_m
    \in \sig$
    is the relation
    \[
    R^{\explic(S)} \defeq
    \midbraces{
      \parens{
        \expel<n_1,d_1>,\dots,
        \expel<n_m,d_m>
      }
    }{
      \begin{array}{l}
        \expel<n_1,d_1> \in \explic(n_1),\\
        \dots,\\
        \expel<n_m,d_m> \in \explic(n_m),\tand\\
        \M*, \sub[d_1/x_1,,d_m/x_m]\\
        \quad\models
        R^S(n_1,\dots,n_m)
      \end{array}
    }
    \]
  \end{citemize}
\end{definition}

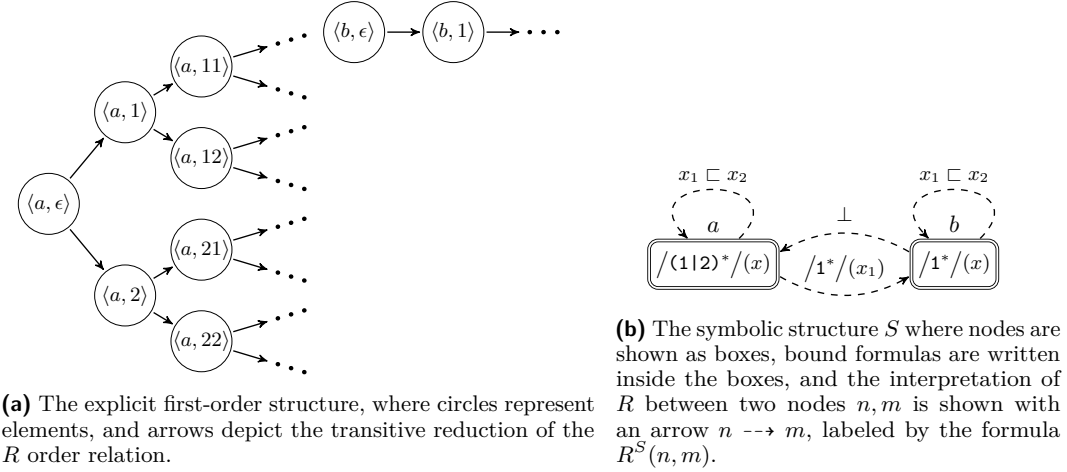
\begin{figure}[t]
\begin{subfigure}{0.56\textwidth}
  \centering
  \scalepic{\begin{tikzpicture}[symbolic,y=0.75cm,scale=0.95]
      \tikzset{
        regular/.append style={
          inner xsep=0pt, 
          minimum size=27pt,
        }
      }
      \node[regular] at (0,0) (a) {$\expel<a,\epsilon>$};
      \node[regular] at (1,2) (a1) {$\expel<a,1>$};
      \node[regular] at (1,-2) (a2) {$\expel<a,2>$};

      \node[regular] at (2,3) (a11) {$\expel<a,11>$};
      \node[] at (3,3.5) (a111dots) {};
      \node[] at (3,2.5) (a112dots) {};

      \node[regular] at (4,3.75) (b) {$\expel<b,\epsilon>$};
      \node[regular] at (5.3,3.75) (b1) {$\expel<b,1>$};
      \node[] at (6.3,3.75) (b1dots) {};

      \node[regular] at (2,1) (a12) {$\expel<a,12>$};
      \node[] at (3,1.5) (a121dots) {};
      \node[] at (3,0.5) (a122dots) {};

      \node[regular] at (2,-1) (a21) {$\expel<a,21>$};
      \node[] at (3,-0.5) (a211dots) {};
      \node[] at (3,-1.5) (a212dots) {};

      \node[regular] at (2,-3) (a22) {$\expel<a,22>$};
      \node[] at (3,-2.5) (a221dots) {};
      \node[] at (3,-3.5) (a222dots) {};

      \path[darrow dots] ($(a111dots)$) -- ($(a11)!1.5!(a111dots)$);
      \path[darrow dots] ($(a112dots)$) -- ($(a11)!1.5!(a112dots)$);
      \path[darrow dots] ($(a121dots)$) -- ($(a12)!1.5!(a121dots)$);
      \path[darrow dots] ($(a122dots)$) -- ($(a12)!1.5!(a122dots)$);
      \path[darrow dots] ($(a211dots)$) -- ($(a21)!1.5!(a211dots)$);
      \path[darrow dots] ($(a212dots)$) -- ($(a21)!1.5!(a212dots)$);
      \path[darrow dots] ($(a221dots)$) -- ($(a22)!1.5!(a221dots)$);
      \path[darrow dots] ($(a222dots)$) -- ($(a22)!1.5!(a222dots)$);
      \path[darrow dots] ($(b1dots)$) -- ($(b1)!1.5!(b1dots)$);

      \path[arrow]
        (a) edge (a1)
        (a) edge (a2)
        (a1) edge (a11)
        (a1) edge (a12)
        (a2) edge (a21)
        (a2) edge (a22)

        (a11) edge (a111dots)
        (a11) edge (a112dots)

        (a12) edge (a121dots)
        (a12) edge (a122dots)

        (a21) edge (a211dots)
        (a21) edge (a212dots)

        (a22) edge (a221dots)
        (a22) edge (a222dots)

        (b) edge (b1)
        (b1) edge (b1dots)
      ;
  \end{tikzpicture}}
  \Description{Circles connected by arrows in a tree.}
  \caption{
    The explicit first-order structure,
    where circles represent elements,
    and arrows depict the transitive reduction of the 
    $R$ order relation.
  }
  \label{fig:symbolic-struct-example-explicit}
\end{subfigure}\hfill \begin{subfigure}{0.42\textwidth}
  \centering
  \scalepic{\begin{tikzpicture}[symbolic]
      \tikzset{
        summary/.append style={inner xsep=3pt}
      }
      \node[summary,label=above:{\large $a$},] 
        at (0,0) (a) 
        { $\nregexp{(1|2)\R*}(x)$ };

      \node[summary,label=above:{\large $b$},] 
        at (3,0) (b) 
        { $\nregexp{1\R*}(x)$ };

      \path[arrow, dashed]
        (a) edge [loop, looseness=5] 
          node [above] {$x_1 \preflt x_2$} (a)
        (b) edge [loop, looseness=5] 
          node [above] {$x_1 \preflt x_2$} (b)
      ;

      \path[arrow, dashed]
        ($(a.east)-(0,5pt)$) to [bend right] 
          node [above,yshift=2pt] {$\nregexp{1\R*}(x_1)$}
          ($(b.west)-(0,5pt)$)
      ;
      \path[arrow, dashed]
        ($(b.west)+(0,5pt)$) to [bend right]
          node [above,yshift=2pt] {$\bot$}
          ($(a.east)+(0,5pt)$)
      ;
  \end{tikzpicture}}
  \Description{Circles connected by arrows in a tree.}
  \caption{
    The symbolic structure $S$
    where nodes are shown as boxes,
    bound formulas are written inside the boxes,
    and the interpretation of $R$ between two nodes
    $n,m$ is shown with an arrow 
    $n \dashrightarrow m$,
    labeled by the formula $R^S(n,m)$.
  }
  \label{fig:symbolic-struct-example-symbolic}
\end{subfigure}
\caption{
  An example of an infinite first-order structure
  for the signature $\braces{R(\cdot,\cdot)}$,
  presented explicitly and symbolically.
  The structure contains infinitely many elements,
  ordered in a prefix order encoded by $R$.
  The elements are of two ``kinds'': $a$-elements and $b$-elements.
  The $a$-elements are ordered in a full, infinite binary tree,
  represented by binary strings over alphabet $1,2$,
  whereas the $b$-elements form an infinite tail for
  the ``top'' branch of $1$-strings of the $a$-binary tree.
}
\label{fig:symstruct-example}
\end{figure} 
\begin{example}
  \label{ex:symbolic-struct}
  Consider the infinite first-order structure
  depicted in \Cref{fig:symbolic-struct-example-explicit}
  over the single-sorted signature $\sig = \braces{R(\cdot,\cdot)}$
  with sort $\sort$ and a binary relation $R$,
  where we have infinitely many elements ordered in a prefix order,
  forming a full binary tree that has an infinite tail ``after''
  one of its branches.
  The figure depicts the transitive reduction of the binary relation $R$
  by arrows.

  \Cref{fig:symbolic-struct-example-symbolic} depicts
  a symbolic structure
  $S$
  over the base theory of $\str_{\braces{1,2}}$
  for $\sig$,
  whose explication is the structure from \Cref{fig:symbolic-struct-example-explicit},
  where the $a$-elements are grouped together into one node
  and the $b$-elements are grouped into a second node.
  Accordingly, the symbolic structure has two nodes
  in its symbolic domain,
  $\domain^S(\sort) = \braces{ a, b }$,
  where $a$ represents an infinite set of binary words,
  $\bound^S(a) = \nregexp{(1|2)\R*}$,
  and $b$ represents an infinite set of unary words,
  $\bound^S(b) = \nregexp{1\R*}$.
  The relation symbol $R$ is defined symbolically as:
  \[
R^S(a,a) \defeq x_1 \preflt x_2,    \quad
  R^S(b,a) \defeq \bot,               \quad
  R^S(a,b) \defeq \nregexp{1\R*}(x_1),\quad
  R^S(b,b) \defeq x_1 \preflt x_2,
  \]
  expressing that for any two $a$-elements or $b$-elements,
  their order (according to $R$) matches the natural prefix order
  on their respective words,
  that no $b$-element comes before an $a$-element,
  and that all $b$-elements come after $a$-elements
  with words matching the $\nregexp{1\R*}$ language.
\end{example}

\begin{theorem}[Model-checking for symbolic structures]
  \label{thm:model-checking}
  Given a symbolic structure $S$
  over base theory $\theory$
  for signature $\sig$
  and sentence
  $\varphi \in \lang_\sig(\emptyset)$,
  there exists a computable sentence
  $\varphi^S \in \lang*(\emptyset)$
  (given by the model-checking transformation
  described in~\cite{infinite-needle})
  such that
  $\explic(S) \models \varphi
  \iff \models* \varphi^S$.
\end{theorem}
\Proof{thm:model-checking}{
  Given a symbolic structure $S$
  and formula $\varphi$,
  we derive $\varphi^S$ by the same transformation
  as in~\cite{infinite-needle}.
  The theorem is proved by induction
  on the structure of $\varphi$,
  following~\cite{infinite-needle}.
}

\begin{remark}
  For symbolic structures that only use
  quantifier-free formulas in bound formulas
  and interpretations,
  the model-checking transformation
  preserves the quantification structure of the original formula.
  Thus, for a base theory $\theory$ that is only decidable for the
  quantifier-free fragment of $\sig*$
  model-checking would still be decidable,
  but only for the quantifier-free fragment of \fol.
\end{remark}

In the sequel, we focus on
symbolic structures over decidable base theories,
which ensures that model-checking is decidable.  \section{The Ordered Self-Cycle (\osc) Family of Fragments}
\label{sec:osc}
In this section we define
the \osc{} family of fragments and state our decidability results.
We start with the definition of
the $\osc[]$ fragments, which are parameterized by a
sentence $\ordax$ that axiomatizes
an order relation. We then present
the various kinds of orders
considered in this paper.
Finally we define the simplified $\osc*[]$ fragments,
which are used in our decidability proofs.

\subsection{Fragments Definition}
\label{sec:osc-definition}

The Ordered Self-Cycle (\osc)
family of first-order logical fragments
is a set of extensions
of the fragment of stratified formulas
(\fsf)~\cite{paxos-made-epr}.
We start by recalling
the properties of \fsf.
The \fsf\ fragment
is the many-sorted variant of the
Effectively PRopositional fragment
(\epr)~\cite{bsr-epr}.
\fsf{} establishes a finite model property
by requiring the quantifier-alternation graph
of each formula in \fsf{} to be acyclic.
The vertices of the quantifier-alternation graph
correspond to sorts in a formula,
and edges correspond
to $\forall\exists$ quantifier alternations,
either explicit in the formula
or implicitly given by function symbols.
In \fsf{}, the finite model property
stems from the fact that
after Skolemization,
there are
only finitely many ground terms,
which allows complete instantiation
of the universal Skolemized formula,
as in \epr.

The \osc\ family of fragments
relaxes the quantifier-alternation restrictions
of \fsf{}
for a single sort,
$\sinf$,
equipped with some axiomatically defined strict order relation,
${\llt} \colon \sinf \times \sinf$.
We denote by $\strict$ the axiomatization of $\llt$
as a strict order relation,
given by the conjunction of the following formulas:
\begin{align*}
  \tag{irreflexivity}
  & \forall x. \neg \parens{x \llt x} \\
  \tag{transitivity}
  & \forall x, y, z. \parens{ x \llt y \land y \llt z }
    \to x \llt z
\end{align*}
Formally, the \osc\ family is defined as follows.

\begin{definition}[\osc]
  The \osc{} family of fragments
  considers first-order signatures
  that include a distinguished sort
  $\sinf$
  and a binary relation symbol
  $\llt \colon \sinf \times \sinf$.
  Given such a signature $\sig$
  and a sentence $\ordax$
  such that $\ordax \models \strict$
  and $\ordax$ is a single-sorted formula
  over the $\sinf$ sort,
  the $\osc[]$ fragment consists of
  sentences of the form
  $\ordax \land \varphi$,
  where:
  \begin{enumerate}
    \item The formula $\varphi$ uses a single variable of sort $\sinf$.
    \item Function and relation symbols
      other than $\llt$ used in $\varphi$
      have at most one argument of sort $\sinf$.
    \item In the quantifier-alternation graph of $\varphi$,
      the only cycles are self-loops at $\sinf$,
      and the only outgoing edges from $\sinf$ are to $\sinf$,
      resulting from function symbols
      whose range is $\sinf$.
    \item All nested function-application terms of sort $\sinf$
      appearing in $\varphi$ are ground.
  \end{enumerate}
\end{definition}

\subsection{Decidability Results for Fragments of \osc}
\label{sec:osc-decidability}
In this section we give an overview
of the \osc{} fragments for which we prove decidability.
First, we consider a strict total order,
denoted $\total$,
axiomatized as
the conjunction of the following formulas:
\begin{align*}
  \tag{irreflexivity}
  & \forall x. \neg \parens{x \llt x} \\
  \tag{transitivity}
  & \forall x, y, z. \parens{ x \llt y \land y \llt z }
    \to x \llt z \\
  \tag{totality}
  &\forall x, y.
    x = y \lor x \llt y \lor y \llt x
\end{align*}

Next, we consider a strict prefix order,
denoted $\pref$,
which is axiomatized as
the conjunction of the following:
\begin{align*}
  \tag{irreflexivity}
  & \forall x. \neg \parens{x \llt x} \\
  \tag{transitivity}
  & \forall x, y, z. \parens{ x \llt y \land y \llt z }
    \to x \llt z \\
  \tag{downards totality}
  &\forall x, y, z.
    \parens{x \llt z \land y \llt z}
    \to \parens{x = y \lor x \llt y \lor y \llt x}
\end{align*}

Finally, we consider two variants
of each of the above orders.
The ``progressive-successor'' variant,
denoted $\psucc$,
where all functions are progressive
and there exists a successor to every element,
and the ``regressive-predecessor'' variant,
denoted $\rpred$,
where all functions are regressive
and there exists a predecessor to every element.
We consider every combination of each of the
$\total$ and $\pref$ axioms,
together with each of the $\psucc$ and $\rpred$ variants.
Formally, we define $\psucc$ to be the conjunction
of
\begin{align*}
  \tag{progressivity}
  & \Land_{f \in \sig} \forall x. x \llt f(x) \\
  \tag{successor existence}
  & \forall x. \exists y. x \llt y
      \land  \forall z.
      x \llt z \to \parens{z = y \lor y \llt z},
\end{align*}
and $\rpred$ to be the conjunction of
\begin{align*}
  \tag{regressivity}
  & \Land_{f \in \sig} \forall x. f(x) \llt x \\
  \tag{predecessor existence}
  & \forall x. \exists y. y \llt x
      \land \forall z.
      z \llt x \to
      \parens{z = y \lor z \llt y}.
\end{align*}

\begin{figure}[t]
  \begin{lstlisting}
  action send(n: node, v: value):
    assume n == root or msg(n, v);
    sent(n, v)       := true;
    msg(left(n), v)  := true;
    msg(right(n), v) := true;
  \end{lstlisting}
  \caption{Simple message broadcast protocol
  for a network of nodes with a binary-tree topology.}
  \label{fig:broadcast-example}
\end{figure} 
\begin{example}
  For an example of the applicability
  of \osc[\pref] let us consider a distributed
  message broadcast protocol,
  in a network with a topology of a binary tree.
  The network has a root node and each node has two children.
  Each node maintains a set of values it received by messages.
  The root node may send any value to its children,
  while any other node may only send values it already received,
  as expressed in~\Cref{fig:broadcast-example}.
  We encode the topology of the network by a prefix order $\llt$,
  axiomatized by $\pref$,
  having \code{root} be minimal according to $\pref$,
  and the functions \code{left} and \code{right},
  encoding the children of a node
  respect the order.
  We further encode the \code{send} action by a formula over 
  two copies of the signature
  that specifies how the \code{sent} and \code{msg} relations
  are modified by the action.
We wish to prove the safety property
  that for any node $n$ that received value $v$,
  there exists an ancestor $n'$ (according to the order in tree topology)
  that sent that value:
  \(
  \forall n \colon \code{node}, v \colon \code{value}.
  \exists n' \colon \code{node}. \code{msg}(n,v) \to
  n' \llt n \land \code{sent}(n', v)
  \).
  This property, along with the encoding of the
  the protocol is in \osc[\pref],
  but not in the more restrictive fragment of \fsf{}.
\end{example}

\begin{theorem}
  \label{thm:decidability}
  The $\osc[]$ fragment is decidable
  for
  \[
  \ordax \in \braces*{
    \total,
    \total\land\psucc,
    \total\land\rpred,
    \pref,
    \pref\land\psucc,
    \pref\land\rpred
  }.
  \]
\end{theorem}

The result for $\osc[\total]$
is a restatement of~\cite{infinite-needle},
while the other 5 results are new.

\subsection{The Simplified \osc* Family of Fragments}
\label{sec:osc-star}
In this section
we define a simplified family of fragments,
denoted \osc*{},
such that satisfiability in \osc{} is reducible
to satisfiability in \osc*{},
a property that our decidability proofs for
fragments of \osc{} build on.
Roughly, formulas in \osc*{} are single-sorted
and only contain simple $c,x,f(x),g(x)$ terms.

\begin{definition}[\osc*]
  The \osc*{} family of fragments considers
  single-sorted first-order signatures
  with a binary relation symbol
  ${\llt}(\cdot,\cdot)$,
  where all other function and relation symbols
  are unary.
  The single sort is denoted $\sinf$.
  Given such a signature $\sig$
  and a sentence $\ordax$ over $\sig$
  such that $\ordax \models \strict$,
  the $\osc*[]$ fragment consists of sentences
  of the form $\ordax \land \varphi$,
  where:
  \begin{enumerate}
    \item $\varphi$ is a positive Boolean combination
    of formulas of the form $\forall x. \phi(x)$,
    where $\phi$ is quantifier-free; and
    \item all terms appearing in $\phi$
    are of the form $c$
    (for some constant symbol $c \in \sig$),
    $x$
    or $f(x)$
    (for some unary $f \in \sig$).
  \end{enumerate}
\end{definition}

Note that $\sig$ may include multiple function symbols, 
all of which are unary and cause self-loops.

\begin{theorem}
  There exists an algorithm that translates
  every formula $\ordax\land\varphi \in \osc[]$
  to some formula
  $\ordax\land\varphi^\star \in\osc*[]$,
  such that there exists a model
  $M \models \ordax\land\varphi$
  iff
  there exists a model
  $M^\star \models \ordax\land\varphi^\star$.
  Moreover,
  $M$ is computable from $M^\star$.
\end{theorem}

The model-preserving translation
leverages key properties of \osc{}.
First, the restrictions of \osc{}
guarantee that after Skolemization,
the signature of \osc{} formulas
can generate only
finitely many ground terms
for sorts other than $\sinf$,
similarly to \fsf{}.
This fact allows us to
Skolemize and then
``instantiate'' formulas in \osc{}
with all ground terms of non-$\sinf$ sorts,
and arrive at an equi-satisfiable universal formula
where the only variables are of sort $\sinf$.
Next, we replace applications
of function and relation symbols
by fresh unary symbols for each tuple of non-$\sinf$
ground terms appearing in them,
considering all possible
equalities between ground terms.
Finally, we perform a sort of flattening
of nested ground terms of $\sinf$,
iteratively replacing each $f(t)$ term
with some fresh constant $c^*$
and adding the axiom
$\forall x. x = t \to f(x) = c^*$
to the formula.

\begin{corollary}
  If $\osc*[]$ is decidable,
  then so is $\osc[]$.
\end{corollary}

\begin{note}
  Since signatures in $\osc*$ are single-sorted,
  we refer to the domains 
  $\domain^M$ of explicit structures
  and the domains 
  $\domain^S$ of symbolic structures
  as sets of elements, respectively nodes,
  i.e., with abuse of notation,
  $\domain^M = \domain^M(\sinf)$
  and
  $\domain^S = \domain^S(\sinf)$.
\end{note}

 \section{Decidability
  via a Symbolic Model Property}
\label{sec:dec}
In this section we prove the various
decidability results of this 
paper~(\Cref{thm:decidability}),
following and generalizing the proof method
of~\cite{infinite-needle}.
We use symbolic structures over decidable base theories
as means to prove decidability
for several fragments of the \osc* family:
we show that every satisfiable formula
in these fragments admits a
satisfying symbolic structure over a decidable base theory.
Since model-checking of such symbolic structures
is decidable,
this \emph{symbolic model property} ensures
that satisfiability is recursively enumerable,
and since unsatisfiability of all \fol{} formulas
is recursively enumerable,
decidability follows.
Formally:

\begin{definition}
  Let $\frag$ be a fragment of \fol{}.
  We say that $\frag$ admits a
  \emph{symbolic model property}
  when there exists a decidable theory $\theory$ such that
  for every satisfiable sentence
  $\psi \in \frag$
  there exists a symbolic structure $S$
  over base theory $\theory$
  such that $\explic(S) \models \psi$.
\end{definition}

\begin{theorem}
  \label{thm:symbolic-model-property}
  Every fragment of \fol{} that admits the
  symbolic model property is decidable.
\end{theorem}
\Proof{thm:symbolic-model-property}{
  Let $\frag$ be a fragment that admits
  a symbolic model property.
  Thus, there exists a semi-decision procedure
  for satisfiability in $\frag$:
  enumerating symbolic structures 
  (by enumerating domain sizes, and terms and formulas in the base theory)
  and then model-checking each structure.
  Since \fol{} has a complete proof system, there exists
  a semi-decision procedure for unsatisfiability
  in $\frag$,
  and therefore $\frag$ is decidable.
}

Next we provide
a general recipe for
establishing a symbolic model property
for \osc*[] fragments (\Cref{sec:proof-recipe}).
In \Cref{sec:dec-total},
we restate the symbolic-model-property proof of
$\osc*[\total]$ using the proof recipe,
and \Cref{sec:dec-pref,sec:dec-pref-variants,sec:dec-total-variants}
give the new decidability results
for 5 additional fragments of $\osc*$ by 
instantiating the general proof recipe 
to prove symbolic model properties for them.

\subsection{Abstract Proof Recipe for \osc* Decidability}
\label{sec:proof-recipe}
The proof of the symbolic model property
for fragments of the \osc* family
works by showing that
given a formula
$\ordax \land \varphi \in \osc*[]$
and a satisfying model
$M \models \ordax \land \varphi$,
a symbolic structure $S$
over a certain decidable base theory $\theory$
can be constructed out of $M$
such that $\explic(S) \models \ordax \land \varphi$.

Fix a model $M \models \ordax \land \varphi$,
the construction of $S$ and its correctness
are derived from the following insight:
since $\osc*[]$ is made up of
positive Boolean combinations of universal formulas
with a single variable $x$,
we can prove that $M$ and $\explic(S)$ are
\osc*-equivalent,
i.e., satisfy the same \osc*[] formulas,
by proving the following
two properties:

\begin{desired}
  \label{desired:order-equivalent}
  If $M \models \ordax$ then
  $\explic(S) \models \ordax$.
\end{desired}

\begin{desired}
  \label{desired:atom-equivalent}
  There exists a surjective function
  $\elcls \colon \domain^M \to \domain^S$
  such that
  for every atomic \osc* formula $\atom (x)$,
  element $d \in \domain^M$,
  and explicit element
  $\expel<\elcls(d),e> \in \explic(\elcls(d))$,
  we have
  $M, \sub[d/x] \models \atom$
  iff
  $\explic(S), \sub[{\expel<\elcls(d),e>}/x] \models \atom$.\footnote{
    Recall that with abuse of notation we denote
    $\domain^M = \domain^M(\sinf)$
    and $\domain^S = \domain^S(\sinf)$.
  }
\end{desired}

\Cref{desired:atom-equivalent}
ensures that every node $n$
in the symbolic domain of $S$ ``mimics''
some explicit element $d$
in the domain of $M$,
in the sense that all the elements in the explication of $n$
satisfy the same atomic formulas as $d$.
In particular, since $M \models \ordax \land \varphi$,
\Cref{desired:atom-equivalent,desired:order-equivalent}
imply that
$\explic(S) \models \ordax \land \varphi$
(by induction on the structure
of $\varphi$).
The construction of $S$ is divided into two parts:
a generic part,
that is shared among the various fragments
of \osc* we consider,
and a specific part that depends on
$\ordax$ and
informs the choice of the base theory.
Similarly,
the proof recipe breaks down the proofs of
\Cref{desired:atom-equivalent,desired:order-equivalent}
into different lemmas,
some of which are proved correct
based on the generic part of the construction alone,
while the rest are proved for each specific setting.
This section lays out the overall proof recipe
and the generic part of the construction,
while the fragment- and base-theory-specific
details are deferred to
\Cref{sec:dec-pref,sec:dec-total,sec:dec-pref-variants,sec:dec-total-variants}.

\subsubsection{Defining the Symbolic Domain}
We define the symbolic domain $\domain^S$
and the surjective function $\elcls$ by
partitioning the domain of $M$
into finitely many equivalence classes,
and set the symbolic domain
to be the finite set of equivalence classes.
To obtain \Cref{desired:atom-equivalent},
we do this partitioning according
to the atomic formulas.
We observe that it suffices to consider atomic formulas over $\sig$
where $x$ appears, to which we refer as \emph{atoms}:

\begin{definition}[Atoms and non-ground terms]
  An \emph{atom} is an atomic formula
  $\atom$ where $x$ appears as a free variable.
  We denote the set of all atoms by
  $\AllAtoms$,
  and the set of non-ground terms appearing in atoms
  by $\mixed \defeq
  \braces{x} \cup \midbraces{f(x)}{f \in \sig}$.
\end{definition}

We say that an element $d \in \domain^M$
satisfies an atom $\atom$
if
$M, \sub[d/x] \models \atom$,
and denote the set of all atoms $d$
satisfies by $\atoms(d)$:
\[
  \atoms(d) \defeq \midbraces{
    \atom \in \AllAtoms
  }{
    M \sub[d/x] \models \atom
  }.
\]

We partition $\domain^M$ into classes
with identical sets of atoms,
defining the equivalence class of an element
$d \in \domain^M$ as
\[
[d] \defeq \midbraces{d' \in \domain^M}{\atoms(d') = \atoms(d)}.
\]
For every class $[d]$ and elements
$d_1, d_2 \in [d]$,
$\atoms(d_1) = \atoms(d_2)$,
and with abuse of notation
we denote those shared atoms by
$\atoms([d])$.

\begin{lemma}
  \label{thm:recipe-partitioning-equiv}
  The above partitioning induces an equivalence relation
  over $\domain^M$,
  and, in particular,
  for every element $d \in \domain^M$
  and atom $\atom \in \AllAtoms$,
  $
    M, \sub[d/x] \models \atom
    \iff
    \atom \in \atoms(d)
    \iff
    \atom \in \atoms([d])
  $.
\end{lemma}
\Proof{thm:recipe-partitioning-equiv}{
  Follows immediately from the definition of
  $\atoms()$.
}

The domain of the symbolic structure $\domain^S$
is defined to be the set of all equivalence classes,
and we use the names `node' and `class'
for the elements of $\domain^S$ interchangeably.
We define the surjective mapping function
$\elcls \colon \domain^M \to \domain^S$
by mapping each element to its equivalence class,
$\elcls(d) \defeq [d]$.

We use the name \emph{regular node}
for any node $n$ where there exists some constant symbol $c$
such that $x = c \in \atoms(n)$,
and otherwise use the name \emph{summary node}.
Intuitively, regular nodes will
be explicated into a single element,
while summary nodes will be
explicated into (infinitely) many elements.

Note that $\domain^S$ is a finite set,
and we define an arbitrary order over its elements,
denoted by $\ll$,
which will be used in the sequel.

Given the above definitions,
we say that an atom
$\atom \in \AllAtoms$
is \emph{observed} by a node
$n \in \domain^S$
if for every explicit element
$\expel<n,e> \in \explic(n)$,
$\explic(S), \sub[{\expel<n,e>}/x] \models \atom
\iff
\atom \in \atoms(n)$.
Note that observance of all atoms
implies the necessary conditions
of \Cref{desired:atom-equivalent},
which can thus be restated
as the following:

\begin{desired}[Atom observance]
  \label{desired:atoms-observance}
  All atoms are observed by all nodes:
  for every node $n$,
  every explicit element
  $\expel<n,e> \in \explic(n)$
  and atom $\atom \in \AllAtoms$,
  $\atom \in \atoms(n) \iff
    \explic(S), \sub[{\expel<n,z>}/x] \models \atom$.
\end{desired}

The rest of the steps of the recipe explain how
to achieve this property
while ensuring that
$\explic(S) \models \ordax$,
where different parts of the construction address different atoms.
Namely, we divide the atoms into
several different \emph{kinds}:
\begin{cdescription}
  \item[Element atoms]
    Atoms containing
    no function applications,
    internally sub-divided into
    \emph{element relation atoms}
    of the form $P(x)$
    where $P \in \sig$ is a unary relation;
    \emph{element equivalence atoms}
    of the form $x = c$
    where $c \in \sig$ is a constant symbol; and
    \emph{element order atoms}
    of the form $x \llt c$ and $c \llt x$.
  \item[Image atoms]
    Atoms of the form
    $\atom \sub[f(x)/x]$,
    where $\atom$ is an element atom,
    and $f \in \sig$
    is a
    unary function symbol.
    For a given function symbol $f$,
    we use the name $f$-image atoms
    for the set of image atoms associated with it.
  \item[Mixed order atoms]
    Atoms of the form $t \llt t', t = t'$,
    where $t, t'$ are non-ground terms in $\mixed$.
\end{cdescription}

Then, for each kind of atom,
a different part of the construction is important.
Some are generic and others specific
We shall see that for element and image atoms,
\Cref{desired:atoms-observance}
can be
proved independently of the order axioms
and base theory in a generic fashion,
while for the mixed order atoms
the proof is more involved,
and hinges on the specifics
of the order and base theory.

\subsubsection{Observing Element Equivalence Atoms
  via Interpretation of Constants
  and Bound Formulas of Regular Nodes}
Recall that nodes $n$
where $x=c \in \atoms(n)$
for some constant symbol $c$ are regular nodes.
The following lemma
ensures that every constant symbol is associated
with a unique regular node.

\begin{lemma}
  \label{thm:one-class-per-ground}
  For every constant symbol $c \in \sig$,
  there exists exactly one regular node $n$
  such that $x=c \in \atoms(n)$.
\end{lemma}
\Proof{thm:one-class-per-ground}{
  Let $d = c^M$, then
  $M, \sub[d/x] \models x=c$
  and by \Cref{thm:recipe-partitioning-equiv},
  $x=c \in \atoms(\elcls(d))$.
  For any other $n' = \elcls(d')$,
  $x = c \in \atoms(n') \iff 
  M, \sub[d'/x] \models x=c
  \iff d' = c^M = d
  \iff n' = \elcls(d)$.
}

Accordingly, for every constant symbol
$c \in \sig$, we define its interpretation
$c^S \defeq \expel<n, \treg>$
where $n$ is the regular node where
$x=c \in \atoms(n)$, 
and $\treg$ is some ground term of the base theory.
The choice of $\treg$ is theory-specific,
and also depends on other theory-specific
properties of the construction,
in particular the interpretations of function symbols.
We provide the specific term
for each fragment in the corresponding sections.

To ensure the well-definedness of constants
we set the bound formula of every regular node $n$
to be $\bound(n) \defeq x = \treg$.

From these definitions follows our first lemma
towards proving
\Cref{desired:atoms-observance}:

\begin{lemma}
  \label{thm:element-equiv-atoms}
  All element equivalence atoms are observed.
\end{lemma}
\Proof{thm:element-equiv-atoms}{
  Let $x=c$ be some element equivalence atom.
  $x=c \in \atoms(n) \iff c^S = \expel<n,\treg>$.
  Since $\bound(n) = x=\treg$,
  it follows that for the single explicit element
  $\expel<n,e> \in \explic(n)$,
  $x=c \in \atoms(n) \iff
  \explic(S), \sub[{\expel<n,e>}/x]
  \models x=c$.
}

\subsubsection{Observing Element Relation Atoms
  via Interpretation of Unary Relations}
Since the atoms inducing the equivalence classes
include atomic formulas of unary relations,
the interpretation of unary relation
symbols is defined straightforwardly.
For every node $n \in \domain^S$
and every unary relation $P \in \sig$,
we define $P^S(n)$ as
\[
  P^S(n) \defeq \begin{cases}
    \top & \tif P(x) \in \atoms(n), \\
    \bot & \tother,
  \end{cases}
\]
and derive the following lemma:

\begin{lemma}
  \label{thm:element-relation-atoms}
  All element relation atoms are observed.
\end{lemma}
\Proof{thm:element-relation-atoms}{
  For any node $n$ and explicit element $\expel<n,e>$,
  $\explic(S),\sub[x/{\expel<n,e>}] \models P(x)
  \iff \models* P^S(n)
  \iff P^S(n) = \top
  \iff P(x) \in \atoms(n)$.
}

\subsubsection{Observing Element Order Atoms
  via Interpretation of \llt{} for Semi-regular Pairs}
For pairs of nodes $n,m$
where at least one of the nodes is regular
(i.e., semi-regular pairs),
we define the interpretation of the order $\llt$
by considering element order atoms,
essentially inheriting the order interpretation
from the order between elements and constants:
\[
  {\llt^S}(n,m) \defeq \begin{cases}
    \top & \text{if
      $x = c \in \atoms(n)$
      and $c \llt x \in \atoms(m)$} \\
    \bot & \text{if
      $x = c \in \atoms(n)$
      and $c \llt x \notin \atoms(m)$} \\
    \top & \text{if
      $x = c \in \atoms(m)$
      and $x \llt c \in \atoms(n)$} \\
    \bot & \text{if
      $x = c \in \atoms(m)$
      and $x \llt c \notin \atoms(n)$} \\
  \end{cases}
\]

The correctness of this definition
is given by the following two lemmas.

\begin{lemma}
  \label{thm:semi-regular-order}
  At least one of the conditions above
  is true for semi-regular pairs,
  and when multiple conditions are true,
  they agree on the definition (\,$\top$ or $\bot$).
\end{lemma}

\Proof{thm:semi-regular-order}{
  Since at least one of $n,m$ is regular,
  one of the above cases must hold.
  Moreover, as stated in \Cref{thm:one-class-per-ground},
  there is exactly one node per constant symbol $c$.
  Let $n,m$ be nodes such that one is regular.
  If $n=m$ then trivially the claim holds.
  Otherwise, let $c_1 \neq c_2$ be regular terms such that
  $x=c_1 \in \atoms(n)$ and $x=c_2 \in \atoms(m)$.
  If $c_1 \llt x \in \atoms(m)$,
  since $x=c_2 \in \atoms(m)$,
  then it must be that $c_1 \llt c_2$ is true in the model $M$,
  and therefore, since $x=c_1 \in \atoms(n)$,
  we have
  $x \llt c_2 \in \atoms(n)$.
  Therefore, when two conditions above hold,
  they must agree.
}

\begin{lemma}
  \label{thm:element-order-atoms}
  All element order atoms are observed.
\end{lemma}
\Proof{thm:element-order-atoms}{
  For any element order atom $x \llt c$
  or $c \llt x$,
  the interpretation of $c$ is 
  $c^S = \expel<n,\treg>$,
  for some regular node $n$.
  Thus, at least one of the cases in the definition
  of the semi-regular order will hold,
  and the atom will be observed appropriately.
}

\subsubsection{Observing Image Atoms
  by Choosing Targets for Function Interpretations}
Recall that in symbolic structures
function interpretations are defined by pairs
of nodes and theory terms.
In particular, for the unary functions
of $\osc*$ we have
$f^S \colon \domain^S \to
\domain^S \times \Terms*(\braces{x_1})$.
When constructing $S$ we can define
the function interpretations in two phases:
first defining the function target node,
in a way that will ensure observance of image atoms,
then using the base theory to define the function terms,
such that the functions are well-defined
and observance of the mixed order atoms is ensured.

Given a node $n$ and function $f$,
we define its interpretation as
$f^S(n) \defeq \expel<m^f_n,t^f_n>$,
where
the target $m^f_n$ is chosen such that all
$f$-image atoms of $n$
match the element atoms of $m$:
\[
  \atom \sub[f(x)/x] \in \atoms(n) 
  \iff \atom \in \atoms(m^f_n).
\]
When multiple candidates exist, we take the least
according to the arbitrary order $\ll$
(to ensure consistency among different definitions).
The following lemma ensures that such node
always exists.

\begin{lemma}
  \label{thm:image-targets}
  For every node $n$ and function symbol $f$,
  there exists some node $m^f_n$ such that
  for every element atom $\atom$,
  $\atom \in \atoms(m^f_n) \iff
  \atom \sub[f(x)/x] \in \atoms(n)$.
\end{lemma}
\Proof{thm:image-targets}{
  Let $n = [d]$ be some node in $S$
  and let $f$ be some function symbol.
  Let us denote by $d' = f^M(d)$ 
  the image of $d$ in $f$ in the model $M$,
  and let us denote the corresponding node by $n' = [d']$.
  For every element atom $\atom$,
  if $\atom \in \atoms(n')$ then,
  by \Cref{thm:recipe-partitioning-equiv},
  $M, \sub[d'/x] \models \atom$
  and since $d' = f^M(d)$ we have
  $M, \sub[d/x] \models \atom \sub[f(x)/x]
  \iff
  \atom \sub[f(x)/x] \in \atoms(n)$.
}

\begin{lemma}
  \label{thm:image-atoms}
  All image atoms are observed.
\end{lemma}
\Proof{thm:image-atoms}{
  Follows from the choice of $m^f_n$.
}

The function terms $t^f_n$ will be defined later,
with the exception that if
$m^f_n$ is a regular node,
then $t^f_n \defeq \treg$,
as dictated by the well-definedness requirement
of symbolic structures.

\subsubsection{Observing Mixed Order Atoms via
  Function Terms,
  Bound Formulas of Summary Nodes
  and Interpretation of \llt{} for Summary Nodes}
\label{sec:mixed-order-recipe}
Finally, the only atoms not yet handled
by the previous steps
are the mixed order atoms,
whose handling is tightly coupled
with the interpretation of the order symbol
and the function terms,
which are in turn
entangled with the definition of
bound formulas for summary nodes.
Beyond the mixed order atoms,
the interpretation of $\llt$
must also satisfy the order axiom $\ordax$.

Thus far, the order interpretation was only defined
for pairs of nodes where at least one is regular,
and only function targets were defined.
Defining the function terms
for function interpretations,
bound formulas of summary nodes
and the interpretation of $\llt$
for summary nodes
is heavily dependent on the specific order axiom $\ordax$
and imposes requirements on the base theory.
In \Cref{sec:dec-total,sec:dec-pref,sec:dec-total-variants,sec:dec-pref-variants}
we complete the details
and present the chosen base theories for the order axioms we consider.

The key idea for defining the remaining components
is partitioning the summary nodes to \emph{segments}
according to their relation
to constant symbols
(i.e., according to the element order atoms).
The exact criteria for segmenting
is dependent on the specific order axiom $\ordax$,
and is chosen to allow us to split
the remainder of the construction and the proof
into two tasks:
\begin{enumerate*}[(i)]
  \item
  \namedlabel{task:inter-segment}{inter-segment task}
  the \emph{inter-segment task};
and
  \item
  \namedlabel{task:intra-segment}{intra-segment task}
  the \emph{intra-segment task}.
\end{enumerate*}

\para{The inter-segment task}
  This task defines 
  the order interpretation between nodes of different segments
  such that mixed order atoms
  involving functions whose
  target nodes reside in different segments
  are observed
  (regardless of the definition of function terms).
  
\para{The intra-segment task and embeddability}
  This task
  defines the order interpretation between nodes
  within the same segment,
  as well as the function terms and bound formulas,
  carefully examining the interaction
  between function terms
  and the order constraints to ensure that
  mixed order atoms
  also hold within a segment
  and that the resulting symbolic structure
  is well-defined.
The base theory is key for this task.
Namely, the base theory is used to define 
\begin{enumerate*}[(i)]
  \item 
  an order interpretation within a summary node,
  which is expanded into an order between any two summary nodes 
  within a segment by interleaving the nodes;
\item function terms
    such that the required order
    on the non-ground terms $\mixed$
    (as induced by the mixed order atoms)
    is maintained for every explicit element;
and
  \item
  bound formulas 
  that ensure that function terms are well-defined, 
  and that the order definition satisfies the order axioms.
\end{enumerate*}
The combination of bound formulas and the order interpretation
determines the \emph{internal shape} of each summary node,
i.e., the ordered set of explicit elements represented by the node.
Defining the function terms in a way that observes the mixed order atoms
requires the internal shape of every summary node 
(which may serve as the target
for some function definition)
to have the property of
\emph{finite embeddability}:
the ability to embed within the node
the set of non-ground terms $\mixed$,
ordered in any way that adheres to the order axioms.
Moreover, to support mixed order atoms involving the term $x$,
we need the stronger property of
\emph{universal finite embeddability}:
the theory terms produced must
be able to embed the order
around every explicit element of the node.
The lifting of the order by interleaving explicit elements
of same-segment nodes
ensures that the same terms 
still comply with the mixed order atoms even when functions target 
different summary nodes within the same segment.
Figuring out how to achieve and encode the
universal finite embeddability property
within the base theory
is one of the key insights required for completing
the proof,
and we shall see how it
dictates the internal shape of summary nodes
in the proofs of~\Cref{sec:dec-total,sec:dec-pref,sec:dec-total-variants,sec:dec-pref-variants}.

\begin{remark}
  As we shall see,
  the proofs given in
  \Cref{sec:dec-total,sec:dec-pref,sec:dec-total-variants,sec:dec-pref-variants}
  actually show a stronger property:
  the fragments of \osc*{} we consider enjoy
  a \emph{bounded} symbolic model property,
  where every satisfiable formula
  has a satisfying symbolic structure
  with bounded size,
  where all
  bound formulas, function terms and relation formulas
  are taken from a finite set of theory terms and formulas,
  dependent only on syntactic properties of the formula.
\end{remark}

 \subsection{Revisiting the Decidability Proof of
  the Totally-Ordered Self-Cycle Fragment}
\label{sec:dec-total}
In this section we restate the proof
of the symbolic model property of $\osc*[\total]$
from~\cite{infinite-needle},
using the more general proof framework we have introduced.
This section is meant as a gentle introduction
to the proof technique,
using the simpler case of a total (linear) order.
The construction for $\osc*[\total]$ uses
the theory of \lia{} as the decidable base theory
for the constructed symbolic structure.

Following the steps given in \Cref{sec:proof-recipe},
we first define the regular node term to be $\treg = 0$,
meaning regular nodes have bound formulas $x = 0$,
function terms are defined $t^f_n \defeq 0$
whenever $m^f_n$ is a regular node,
and
each constant $c$ is interpreted as
$\expel<n,0>$
where $n$ is the regular node
for which
${x = c} \in \atoms(n)$.

Next we define the segments of summary nodes
in a way that lets us easily resolve the \nameref{task:inter-segment},
namely, define the inter-segment order interpretation.
For every summary node $n$
we denote the set of constant symbols
``smaller'' than $n$
by
$
\segS(n) \defeq
\midbraces{ c \in \sig }{ c \llt x \in \atoms(n)}
$,
and we define a segment of summary nodes
as a set of nodes with identical
$\segS()$.
Using $\segS()$ we define
the inter-segment order as follows.
Given two nodes $n,m$ with segments
$\segS(n) \neq \segS(m)$,
\[
  {\llt^S}(n,m) \defeq \begin{cases}
    \top &\tif \segS(n) \subsetneq \segS(m), \\
    \bot &\tother. \\
  \end{cases}
\]

This definition of the inter-segment order interpretation
is dictated by the total order axiom $\total$.
Intuitively, if there exists some constant symbol $c$
such that $c \llt x \in \atoms(m)$
and $c \llt x \notin \atoms(n)$
then by totality,
$x \llt c \in \atoms(n)$
($n$ is a summary node, thus $x = c \notin \atoms(n)$),
and by transitivity,
it must be the case that
all explicit elements of $n$ are smaller
than those of $m$.
Note that given that $\segS(n) \neq \segS(m)$,
the total order axiom ensures that if
$\segS(n) \subsetneq \segS(m)$ does not hold,
then the other direction must hold, i.e.,
$\segS(m) \subsetneq \segS(n)$.

We have yet to define
the interpretation of $\llt$
for same-segment summary nodes,
thereby completing the definition of $\llt$.
Yet, remarkably,
due to the careful definition of the segments
we can prove
the correctness of the construction
regarding the order axiom in a modular fashion,
before defining the intra-segment order.

\begin{lemma}
  \label{thm:tot-global-order}
  For any definition of the intra-segment order
  that produces a strict total order
  within each segment,
  the interpretation $\llt^{\explic(S)}$ is a total order,
  i.e., $\explic(S) \models \total$.
\end{lemma}
\Proof{thm:tot-global-order}{
  We prove irreflexivity, transitivity and totality separately.
  Proof by simple case analysis.
}

Moreover, the inter-segment order definition
ensures observance of mixed order atoms
involving nodes of different segments:

\begin{lemma}
  \label{thm:tot-inter-segment-mixed-atoms}
  Let $n$ be a summary node
  and let $f,g \in \sig$
  be function symbols such that
  $m^f_n$ and $m^f_n$ reside within different segments,
  then the mixed order atoms
  $f(x) \llt g(x)$ and $g(x) \llt f(x)$ are observed.
  Further, for $f$ such that $m^f_n$ and $n$ reside
  in different segments,
  $x \llt f(x)$ and $f(x) \llt x$ are observed.
\end{lemma}
\Proof{thm:tot-inter-segment-mixed-atoms}{
  Follows from the definition of the inter-segment order.
}

Next we approach the \nameref{task:intra-segment}.
Namely, we define
the intra-segment order,
function terms and bound formulas
for summary nodes, in a way that observes all mixed order atoms.
As explained in \Cref{sec:proof-recipe},
this is achieved by the universal embeddability property,
which ensures that
for any possible set of constraints
over the non-ground terms,
i.e., any possible set of mixed order atoms
in some summary node,
these constraints can be satisfied
\emph{for every explicit element} associated with
the summary node.

\begin{figure}
\begin{minipage}[b]{0.49\textwidth}
\centering
\scalepic{\begin{tikzpicture}[symbolic]
\node at (0.85,0) (predots) {$\cdots <\;\, $};
    \node[explicit] at (1.5,0) (fx) {$f(x)$};
    \node at (2,0) {$<$};
    \node[explicit] at (2.5,0) (gx) {$g(x)$};
    \node at (3,0) {$<$};
    \node[explicit] at (3.5,0) (x) {$x$};
    \node at (4,0) {$<$};
    \node[explicit] at (4.5,0) (x) {$h(x)$};
    \node at (5.15,0) (postdots) {$\;\,< \cdots$};
    \node[summaryFit, 
      fit=(predots) (x) (postdots), 
      inner xsep=2pt,
      label=above:{\large $n$},
    ] {};
\end{tikzpicture}}
\Description{Circles connected by arrows in a line, 
  with two squares on each edge.}
\caption{
  Example of an order induced by mixed order atoms
  within a single summary node.}
\label{fig:tot-segment-example} \end{minipage}\hfill \begin{minipage}[b]{0.49\textwidth}
\centering
\begingroup
\setlength{\nodeSize}{24pt}
\scalepic{\begin{tikzpicture}[symbolic]
  \node[regular] at (-0.5,0) (c1) {$c_1$};
  \node at (0.25,0) (predots) {\huge $\cdots$};
  \node[explicit] at (1,1) (n1-0) {$\expel<n_1,0>$};
  \node[explicit] at (1.5,0) (n2-0) {$\expel<n_2,0>$};
  \node[explicit] at (2,-1) (n3-0) {$\expel<n_3,0>$};
  \node[explicit] at (2.5,1) (n1-1) {$\expel<n_1,1>$};
  \node[explicit] at (3,0) (n2-1) {$\expel<n_2,1>$};
  \node[explicit] at (3.5,-1) (n3-1) {$\expel<n_3,1>$};
  \node at (4.25,0) (postdots) {\huge $\cdots$};
  \node[regular] at (5,0) (c2) {$c_2$};

  \path[arrow] 
    (predots.north east) -- (n1-0)
    (n1-0) edge (n2-0)
    (n2-0) edge (n3-0)
    (n3-0) edge (n1-1)
    (n1-1) edge (n2-1)
    (n2-1) edge (n3-1)
    (n3-1) -- (postdots.south west)
  ;

  \draw[thick]
    ([yshift=30pt] c1.north) -- ([yshift=1pt] c1.north)
    ([yshift=-1pt] c1.south) -- ([yshift=-30pt] c1.south);
  \draw[thick]
    ([yshift=30pt] c2.north) -- ([yshift=1pt] c2.north)
    ([yshift=-1pt] c2.south) -- ([yshift=-30pt] c2.south);
\end{tikzpicture}}
\endgroup
\Description{Circles connected by arrows in a line, 
  with dots on each side.}
\caption{
  A segment with three interleaving summary nodes.}
\label{fig:tot-interleaving} \end{minipage}
\end{figure}

For example, consider a summary node $n$
where the image atoms of $n$ constrain
the functions $f,g$ and $h$ to map back to $n$
(i.e., $n$ is the function target for $f,g$ and $h$),
and the mixed order atoms of $n$
induce the following order:
$f(x) \llt g(x) \llt x \llt h(x)$
(see \Cref{fig:tot-segment-example}).
We can see that for any explicit element of $n$,
there must be two smaller explicit elements of $n$,
which $f(x)$ and $g(x)$ will map to,
and one larger explicit element of $n$,
which $h(x)$ will map to.
Further note that all of these explicit elements
must be linearly ordered.

One natural shape for such a summary node
is a bidirectionally-infinite line ---
the integers ordered by $<$.
Notably, for the case of $\osc*[\total]$,
this shape is uniform for all summary nodes.
(We shall see that for prefix orders,
two internal shapes are required).
Using the base theory of \lia{},
we define the bound formulas
for all summary nodes as simply
$\bound(n) = \top$.

Back in our example,
we can now define the function terms
for $f,g$ and $h$
by considering the arrangement
of the terms $f(x), g(x), h(x)$
relative to $x$,
according to the mixed order atoms.
We choose function terms that realize
this arrangement,
for example,
$t^f_n = x_1 - 2,
t^g_n = x_1 - 1$
and $t^h_n = x_1 + 1$.

The discussion so far considered a single summary node. 
As explained earlier (\Cref{sec:mixed-order-recipe}), 
when a segment includes multiple summary nodes, 
we lift the order to nodes within the same segment by interleaving 
the explicit elements of the different summary nodes within the segment
(which ensure function term compliance with mixed order atoms 
across different target summary nodes within the segment). 
An example is given in \Cref{fig:tot-interleaving}.
The interleaving order is unimportant but should be consistent,
and we use the arbitrary order $\ll$
to deterministically define it.

Finally, we provide the formal definition of the intra-segment order 
and the function terms.
We define the intra-segment order between
two nodes $n,m$ as follows:
\[
  n \llt^S m \defeq \begin{cases}
    x_1 < x_2 \lor x_1 = x_2 & \tif n \ll m, \\
    x_1 < x_2 & \tother.
  \end{cases}
\]
To define the function terms
for the cases where the target node
of the function interpretation is a summary node (as illustrated above),
we consider for every node its set of
mixed order atoms.
We order all non-ground terms $\mixed$
linearly, according to the order induced
by the mixed order atoms.
Note that two terms $t,t'$
may occupy the same ``spot'' in the order,
if $t = t' \in \atoms(n)$
(e.g., $f(x) = g(x) \in \atoms(n)$).
Using the non-ground term $x$ as an anchor,
for each non-ground term $f(x)$
we denote by $k^f_n$
the distance between it and the anchor term~$x$,
and define
$t^f_n \defeq x_1 + k^f_n$.

We formalize
the correctness of the construction
as follows:

\begin{lemma}
  \label{thm:tot-intra-segment-order}
  The intra-segment definition of the
  interpretation of $\llt$
  produces a strict total order
  within each segment.
\end{lemma}
\Proof{thm:tot-intra-segment-order}{
  Follows from the properties of the $<$ relation
  on the integers.
}

\begin{lemma}
  \label{thm:tot-well-defined}
  The constructed symbolic structure
  is well-defined.
\end{lemma}
\Proof{thm:tot-well-defined}{
  Function terms for regular target nodes are
  always $\treg$,
  which satisfies the bound formulas
  $x=\treg$.
  The bound formulas for summary nodes are $\top$,
  which are always satisfied.
}

\begin{lemma}
  \label{thm:tot-mixed-atoms}
  All mixed order atoms are observed.
\end{lemma}
\Proof{thm:tot-mixed-atoms}{
  Proof by case analysis
  of non-ground terms and their segments.
}

This finishes the proof of
\Cref{desired:atoms-observance}.
\Cref{thm:tot-global-order,thm:tot-intra-segment-order}
imply that $\explic(S) \models \total$,
and therefore:

\begin{corollary}
  $\explic(S) \models \total \land \varphi$.
\end{corollary}

 \subsection{Decidability of the
  Prefix-Ordered Self-Cycle Fragment}
\label{sec:dec-pref}

In this section we prove the main result of this paper:
the decidability of the $\osc[\pref]$ fragment.
To that end, we prove the symbolic model property for $\osc*[\pref]$
using the $\str$ base theory.
As before, we follow the steps of \Cref{sec:proof-recipe}.
However, $\osc*[\pref]$ requires more involved
handling by the base theory at every step.

First, we fix the base theory to
be $\str_\Delta$ over alphabet
$\Delta = \braces{ 0, \dots, \ell }$,
where $\ell = \abs{\mixed}$, 
i.e., $\ell$ is 
the number of non-ground terms
$x, f(x), g(x), \ldots$ that appear in atoms
(and assumed to be at least 2).
Intuitively, $\ell$ letters in the alphabet 
are sufficient
to accommodate $\ell$
non-ground terms that may be incomparable (for some element),
and the letter $0$ is used to represent elements that are smaller than all of them.

The regular-node term
is the empty string
$\treg = \epsilon$,
thereby defining
bound formulas for regular nodes,
interpretations of constants,
and function terms for regular-node targets.

Next we define the segments of summary nodes.
Whereas in the total order,
each segment represented some interval
after a subset of the linearly ordered constants,
for a prefix order,
a finite set of constants may be ordered
as a forest of rooted trees.
Moreover, the space between a constant
and its successor constants in a tree
may itself be structured as a tree,
inducing multiple segments.

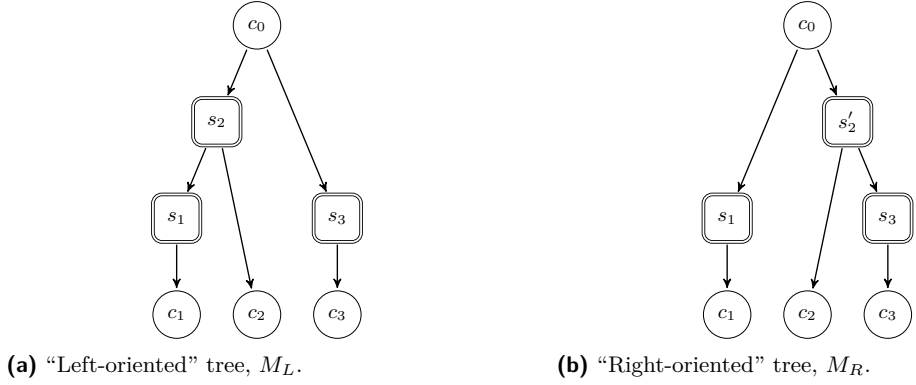
\begin{figure}
  \begin{subfigure}{0.48\textwidth}
  \centering
  \scalepic{\begin{tikzpicture}[symbolic]
      \node[regular] at (0,3) (c0) {$c_0$};
      \node[regular] at (-1,-1.5) (c1) {$c_1$};
      \node[regular] at (0,-1.5) (c2) {$c_2$};
      \node[regular] at (1,-1.5) (c3) {$c_3$};

      \node[summary] at (-1,0) (s1) {$s_1$};
      \node[summary] at (-0.5,1.5) (s2) {$s_2$};
      \node[summary] at (1,0) (s3) {$s_3$};

      \path[arrow] 
        (c0) edge (s2)
        (c0) edge (s3)
        (s1) edge (c1)
        (s2) edge (c2)
        (s2) edge (s1)
        (s3) edge (c3)
      ;
  \end{tikzpicture}}
  \Description{Circles connected by arrows in a tree.}
  \caption{
    ``Left-oriented'' tree, $M_L$.
  }
  \label{fig:prefix-segments1}
  \end{subfigure}\hfill \begin{subfigure}{0.48\textwidth}
  \centering
  \scalepic{\begin{tikzpicture}[symbolic]
      \node[regular] at (0,3) (c0) {$c_0$};
      \node[regular] at (-1,-1.5) (c1) {$c_1$};
      \node[regular] at (0,-1.5) (c2) {$c_2$};
      \node[regular] at (1,-1.5) (c3) {$c_3$};

      \node[summary] at (-1,0) (s1) {$s_1$};
      \node[summary] at (0.5,1.5) (s2) {$s'_2$};
      \node[summary] at (1,0) (s3) {$s_3$};

      \path[arrow] 
        (c0) edge (s2)
        (c0) edge (s1)
        (s1) edge (c1)
        (s2) edge (c2)
        (s2) edge (s3)
        (s3) edge (c3)
      ;
  \end{tikzpicture}}
  \Description{Circles and rectangles 
    connected by arrows in a tree.}
  \caption{
    ``Right-oriented'' tree, $M_R$.
  }
  \label{fig:prefix-segments2}
  \end{subfigure}
  \caption{
    Two possible sets of segments
    laid over 4 constants
    ordered in a tree.
    The arrows depict the transitive reduction.
  }
\end{figure} 
\begin{example}
  \label{ex:prefix-segments}
  \Cref{fig:prefix-segments1,fig:prefix-segments2}
  depict two possible sets of segments
  between 4 constant symbols that
  are ordered in a tree.
  We denote by $M_L,M_R$ two structures
  for a signature $\braces{c_0,c_1,c_2,c_3}$,
  where for both we have
  $M_* \models
  c_0 \llt c_1 \land c_0 \llt c_2 \land c_0 \llt c_3$
  and for any other pairs $c_i,c_j$,
  $M_* \not\models c_i \llt c_j$.
  However, $M_L$ and $M_R$ induce different segments,
  as in $M_L$ there exists an element
  that is smaller than $c_1$ and $c_2$
  but incomparable to $c_3$,
  and in $M_R$ there exists an element that is
  smaller than $c_2$ and $c_3$
  but incomparable to $c_1$.
\end{example}

Formally, we define the segments
in the construction
as pairs of sets of constant symbols,
recording for each node $n$
both the set of constant symbols
smaller than $n$,
$\segS(n) \defeq \midbraces*{
  c \in \sig
}{
  c \llt x \in \atoms(n)
}$ (as in $\osc*[\total]$),
as well as the set of constant symbols
larger than $n$,
$\segL(n) \defeq \midbraces*{
  c \in \sig
}{
  x \llt c \in \atoms(n)
}$.
We denote the segment of a node $n$ by
$\seg(n) \defeq \parens{ \segS(n), \segL(n) }$.

Having defined the segments of summary nodes,
we now address the \nameref{task:inter-segment}.
We start with an example.

\begin{example}
  Going back to \Cref{ex:prefix-segments},
  let us focus on $M_L$ of
  \Cref{fig:prefix-segments1}.
Formally, we have 3 summary nodes $s_1, s_2, s_3$
  where
  $\seg(s_1) = \parens{ \braces{c_0}, \braces{c_1} }$,
  $\seg(s_2) = \parens{ \braces{c_0}, \braces{c_1,c_2} }$
  and
  $\seg(s_3) = \parens{ \braces{c_0}, \braces{c_3} }$.
The element order atoms imply that
  $s_1$ is incomparable to $s_3$,
  as otherwise, by transitivity 
  we would expect $x \llt c_3 \in \atoms(s_1)$
  or $x \llt c_1 \in \atoms(s_3)$.
  Similarly, $s_2$ and $s_3$ are incomparable.
  In contrast, since both $s_1$ and $s_2$
  include the $x \llt c_1$ atom,
  by downwards totality
  they must be ordered.
  Moreover, 
  $s_2$ must come before $s_1$,
  since otherwise, by transitivity,
  we would expect $x \llt c_2 \in \atoms(s_1)$.
\neta{(OLD) the example does not show
  the first case of the inter-segement definition}
\end{example}

The intuitive inter-segment order
derived in \Cref{ex:prefix-segments}
is generalized and formally defined
in the following way.
Given two nodes $n,m$ such that $\seg(n) \neq \seg(m)$:
\[
  {\llt^S}(n,m) \defeq \begin{cases}
    \top &\tif \segL(n) \cap \segS(m) \neq \emptyset, \\
    \top &\tif \segS(n) = \segS(m) \tand \segL(m) \subsetneq \segL(n),\\
    \bot &\tother.
  \end{cases}
\]

\begin{remark}
  Interestingly,
  these same definitions
  also work
  for $\osc*[\total]$.
  Though redundant,
  we could have defined the segments
  for $\osc*[\total]$ to include both
  $\segS()$ and $\segL()$,
  and use the above definition
  for the inter-segment order.
  Note that the first case is equivalent to
  the first case in the definition of the
  inter-segment order for $\osc*[\total]$,
  and the second case is void in $\osc*[\total]$.
  In a total order,
  if $\segS(n) = \segS(m)$ then
  $\segL(n) = \segL(m)$
  (and in particular
  it cannot be that $\segL(m) \subsetneq \segL(n)$).
  This property can be understood intuitively,
  since a total order is a special case of a prefix order.
\end{remark}

Similarly to the construction for $\osc*[\total]$,
the definition of the inter-segment order
obtains the following lemmas:

\begin{lemma}
  \label{thm:pref-global-order}
  For any definition of the intra-segment order
  that produces a strict prefix order
  within each segment,
  the interpretation $\llt^{\explic(S)}$
  is a strict prefix order,
  i.e., $\explic(S) \models \pref$.
\end{lemma}
\Proof{thm:pref-global-order}{
  Similarly to \Cref{thm:tot-global-order},
  proving irreflexivity, transitivity and
  downwards-totality separately,
  by case splitting.
}
\sharon{consider formalizing. If for every segment $A$,
$\llt^{\explic(S)} \cap (\bigcup_{n\in A} \explic(n) \times \bigcup_{n\in A} \explic(n)$ is a strict prefix order, then
the interpretation $\llt^{\explic(S)}$
  is a strict prefix order,
  i.e., $\explic(S) \models \pref$. }

\begin{lemma}
  \label{thm:pref-inter-segment-mixed-atoms}
  Let $n$ be a summary node
  and let $f,g \in \sig$
  be function symbols such that
  $m^f_n$ and $m^f_n$ reside within different segments,
  then the mixed order atoms
  $f(x) \llt g(x)$ and $g(x) \llt f(x)$ are observed.
  Further, for $f$ such that $m^f_n$ and $n$ reside
  in different segments,
  $x \llt f(x)$ and $f(x) \llt x$ are observed.
\end{lemma}
\Proof{thm:pref-inter-segment-mixed-atoms}{
  Follows from the definition of the inter-segment order.
}

Moving on to the \nameref{task:intra-segment},
we first explain the internal shape of summary nodes, which is defined by the combination of bound formulas
and order interpretation.
We distinguish two kinds of summary nodes:
linear nodes,
any node $n$ where $\segL(n) \neq \emptyset$,
and spanning nodes,
where $\segL(n) = \emptyset$.
Note that when $\segL(n) \neq \emptyset$,
there exists some constant $c$ such that
$n$ has atom
$x \llt c \in \atoms(n)$,
thus all explicit elements of $n$ must be
smaller than $c$, and therefore,
by downwards totality,
totally ordered.

The internal shape of a linear node $n$ is isomorphic
to the integer number line,
which we express using a unary encoding:
a non-negative integer $z$ is represented by the string
$1^z$ (i.e., concatenating the letter `$1$' $z$ times),
and a negative integer $z$ is represented by the string
$0^{-z}$.
The resulting bound formula
$\bound(n)$
is given by the regular expression
$\nregexp{0\R+|1\R*}(x)$
(recall that any regular expression is
a relation symbol in the signature of the \str{} theory).

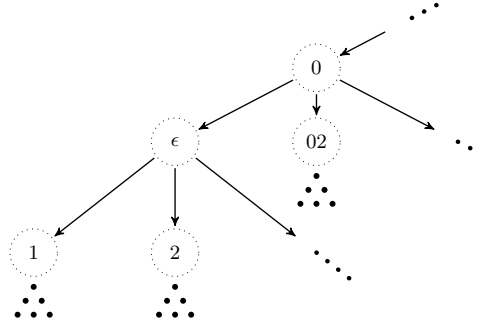
\begin{figure}
\centering
\scalepic{\begin{tikzpicture}[symbolic,x=2.2cm,y=1.15cm]
    \node[dummy] (dots) at (1,1) {};

    \node[explicit] (n0)    at (0,0) {$0$};

    \node[explicit] (n) at (-1,-1) {$\epsilon$};
    \node[explicit] (n02) at (0,-1) {$02$};
    \pic 							    at (0,-1-0.1)  {tridots};
    \node[dummy] (n0-dots) at (1,-1) {};

    \node[explicit] (n1) at (-2,-2-0.5) {$1$};
    \pic at (-2,-2-0.5-0.1)  {tridots};
    \node[explicit] (n2) at (-1,-2-0.5) {$2$};
    \pic 							      at (-1,-2-0.5-0.1)  {tridots};
    \node[dummy] (n-dots) at (0,-2-0.5) {};

    \path[darrow dots] (dots) -- ($(dots)!0.4!(n0)$);
    \path[arrow] ($(dots)!0.5!(n0)$) -- (n0);

    \path[darrow dots] ($(n0-dots)$) -- ($(n0)!1.25!(n0-dots)$);
    \path[darrow dots] ($(n-dots)$) -- ($(n)!1.25!(n-dots)$);

    \path[arrow]
        (n0) edge (n)
        (n0) edge (n02)
        (n0) edge (n0-dots)

        (n) edge (n1)
        (n) edge (n2)
        (n) edge (n-dots)
      ;
\end{tikzpicture}}
\Description{Circles connected by arrows 
  in a bidirectionally-infinite tree.}
\caption{
  The universal tree, a bidirectionally-infinite tree.
  The arrows show the transitive reduction
  of the order of the vertices.
  The ${\protect\tridots}$ symbol
  represents an infinite \emph{rooted} tree,
  isomorphic to the tree represented by the language
  $\nregexp{(1\R-$\ell$)\R*}$
  with the standard prefix order,
  appended to the string of the node it is attached to.
}
\label{fig:prefix-universal-tree}
\end{figure} 
For a spanning node $m$, a richer internal shape
must be supported
to ensure universal finite embeddability
of any prefix order over $\ell$ elements, 
where $\ell$ is the number of non-ground terms in $\mixed$,
thereby facilitating observance of any combination of mixed order atoms.
The general shape is shown in \Cref{fig:prefix-universal-tree},
a bidirectionally-infinite tree
with an out-degree of $\ell$.
It can be seen as an extension
of the infinite line of linear segments,
where each point becomes a root of an infinite tree.
Each (rooted) infinite tree with out-degree $\ell$
can be represented by the regular expressions
$\nregexp{(1\R-$\ell$)\R*}$,
and the entire universal tree is given by the bound formula
$\bound(m) =
\nregexp{(1\R-$\ell$)\R*|(0\R+\R.((2\R-$\ell$)\R.(1\R-$\ell$)\R*)?)}(x)$.

Having defined the bound formulas,
it remains to define the order interpretation
in a way that will induce the desired
internal shape for each summary node.

\subsubsection{Defining the Intra-segment Order by Encoding the Order
  of the Universal Tree}
\label{sec:dec-pref-order}
Order between two vertices in the universal tree,
represented by the words $x_1, x_2$
is given by the formula
$\beta(x_1,x_2)$,
defined in the sequel.
Conveniently, linear summary nodes
are a special case of the universal tree
(where the out-degree is $1$ instead of $\ell$),
and so the same $\beta(x_1,x_2)$
is also suitable for them.

As before, the intra-segment order is given
by interleaving the summary nodes of each segment
arbitrarily
(using $\ll$),
so for nodes $n,m$ where $\seg(n) = \seg(m)$
we define
\[
{\llt^S}(n,m) \defeq \begin{cases}
  \beta(x_1, x_2) \lor x_1 = x_2 &\tif n \ll m,\\
  \beta(x_1,x_2) &\tother.
\end{cases}
\]

In order to derive the order formula $\beta(x_1,x_2)$
between two vertices in the universal tree,
we identify the three different parts
of the universal tree,
shown in \Cref{fig:prefix-universal-tree},
and then define the order within and across each part.
We start by identifying the three different parts:
\begin{enumerate*}[(i)]
  \item 
  the positive tree,
  the rooted, infinitely increasing tree,
  represented by all strings over
  sub-alphabet $\braces{1, \dots, \ell}$,
  \item
  the negative spine,
  the infinitely decreasing line,
  represented by the unary non-empty strings
  $0, 00, \dots$,
  and 
  \item
  the negative branches,
  all trees that span out from vertices 
  on the negative spine.
\end{enumerate*}
These three parts are captured by the following
regular expressions:
\begin{align*}
  \treePos(x) &\defeq
    \nregexp{(1\R-$\ell$)\R*}(x) \\
  \treeNegSpine(x) &\defeq
    \nregexp{0\R+}(x) \\
  \treeNegBranch(x) &\defeq
    \nregexp{0\R+\R.(2\R-$\ell$)\R.(1\R-$\ell$)\R*}(x)
\end{align*}

We now consider all possible pairs of words $x_1,x_2$
such that $x_1$ should come before $x_2$
according to the order within the universal tree.
Immediately we can see that
all vertices in the negative spine
are smaller than vertices in the positive tree.
Two other simple cases are when both words are
in the positive tree or both are in the negative spine:
in the first case they are ordered according to the prefix order,
in the second according to the reverse prefix order.

Lastly, we consider the cases where
the second word $x_2$
is on
the negative branches.
In these cases we need to consider
the vertex on the negative spine from which
the negative branch of $x_2$ is spanning out.
We identify the root of the negative branch
for a word $x$
with the term
$\treeNegRoot(x) \defeq \prefOf_0(x)$,
where $\prefOf_0(x)$
is the longest prefix of $x$ containing 
only `$0$'s. This function is definable in $\str$,
by, e.g., the following formula
$\forall x, y.
y = \prefOf_0(x) \liff
\parens{
y \preflt x \land \nregexp{0\R*}(y) \land
\neg \parens{ y \cdot 0 \preflt x }
}$.
If the first word $x_1$ is 
on the negative spine and comes 
before the root of $x_2$, 
then $x_1$ is smaller than $x_2$.
If $x_1$ is on the same negative branch 
as $x_2$,
then $x_1$ and $x_2$ are compared according 
to the regular prefix order.
In all other cases, $x_1$ and $x_2$
are incomparable.

Putting all of the above together,
$\beta(x_1,x_2)$ is defined as the following
if-then-else expression:
\[
\beta(x_1,x_2) \defeq \begin{cases}
  \top
    &\tif \treeNegSpine(x_1) \land \treePos(x_2) \\
  x_1 \preflt x_2
    &\tif \treePos(x_1) \land \treePos(x_2) \\
  x_2 \preflt x_1
    &\tif \treeNegSpine(x_1) \land \treeNegSpine(x_2) \\
  \treeNegRoot(x_2) \preflt x_1
    &\tif \treeNegSpine(x_1) \land \treeNegBranch(x_2) \\
  x_1 \preflt x_2
    &\tif \treeNegBranch(x_1) \land \treeNegBranch(x_2) 
    \land \treeNegRoot(x_1) = \treeNegRoot(x_2) \\
  \bot &\tother
\end{cases}
\]

\todo{
  examples for $\beta$?
}

\begin{lemma}
  \label{thm:pref-intra-segment-order}
  The intra-segment definition of the
  interpretation of $\llt$
  produces a strict prefix order
  within each segment.
\end{lemma}
\Proof{thm:pref-intra-segment-order}{
  Case analysis of all possible ``locations''
  in the infinite line (for linear nodes)
  or the universal tree (for spanning nodes).
}

\subsubsection{Defining Function Terms by Encoding Finite Paths in the Universal Tree}
\label{sec:dec-pref-terms}
Finally we explain how to define function terms
in a way that respects the mixed order atoms.
Unlike the uniform segments 
used in the construction for $\osc*[\total]$,
the segments for $\osc*[\pref]$ are of two kinds:
linear and spanning, with different internal shapes of nodes.
(Note that all nodes in a segment are of the same kind;
we thus refer to segments as linear or spanning accordingly.)
While in $\osc*[\total]$, the uniformity of segments
allows defining all function terms without regard to the target nodes,
a subtlety arises in $\osc*[\pref]$,
since linear target nodes 
can only embed linear (total) orders, 
which means that we may fail to embed 
the set of \emph{all} non-ground terms in $\mixed$,
ordered according to the mixed order atoms,
in linear nodes.

Therefore, when defining the function terms of a node $n$,
instead of considering all function interpretations
and mixed order atoms together,
we first group functions according to the segments
of their target nodes. 
Each such ``function group''
is associated with a unique segment of the corresponding target nodes,
where the corresponding non-ground terms need to be embedded.
We identify each group
with the set of corresponding non-ground terms,
and when $n$ resides within the group's segment,
we also include $x$.

For example, consider a node $n$ and two function symbols $f,g$ where
the target node of $f$, $m^f_n$, is within the same segment as $n$, and  the target node of $g$, $m^g_n$, is in a different segment.
Then $f$ and $g$ have distinct function groups,
where the function group of $f$ includes the terms $x,f(x)$
and is associated with a segment that includes both $n$ and $m^f_n$, and the function group of $g$ includes only $g(x)$ and is associated with a distinct segment that includes $m^g_n$.

For every function group 
$G \subseteq \mixed$
of non-ground terms,
the function terms for all function symbols
appearing in $G$ are defined jointly,
which is crucial to ensure 
the correct embedding of $G$,
ordered according to the mixed order atoms.
Importantly, the partitioning of non-ground terms $\mixed$
into groups of same-segment target nodes
guarantees that the induced order
can always be embedded in the corresponding segment.

\begin{lemma}
  \label{thm:pref-group-induced-order}
  Given a node $n \in \domain^S$
  and function group of non-ground terms 
  $G \subseteq \mixed$,
if the mixed order atoms of $n$
  induce a non-linear order on $G$,
  then the segment of $G$ is spanning.
\end{lemma}

Generalizing the total order case,
we define the function terms 
for the functions in each group
relative to an anchor;
if $x$ is included in the group
then it is the anchor,
otherwise an arbitrary term $f(x)$
within the group is used.
Instead of using the distance between
a non-ground term $f(x)$ and the anchor term
to produce function terms 
(as was possible in $\osc*[\total]$),
we now think of the prefix order on a group's non-ground terms as a forest,
and consider paths from the anchor as a way to define the function terms.
The paths are composed
from ``parent'', ``child'' and ``sibling'' steps,
where different trees in the forest
are connected by considering their roots as siblings.
In the sequel we describe how to derive
function terms according to the paths
in this forest of non-ground terms.

The paths in the forest of non-ground terms are finite,
and since there are at most
$\ell$ vertices in the forest,
the paths can be encoded
by composing terms for
the parent of a vertex,
$\treeParent(x)$,
the j\textsuperscript{th} child of a vertex,
$\treeChild_j(x)$
(for $1 \leq j \leq \ell$),
and the j\textsuperscript{th} sibling,
$\treeSibling_j(x)$.

For $\treeParent(x)$ we need to consider two cases:
either $x$ is on the non-positive spine,
in which case its parent is obtained by
appending $0$,
otherwise, its parent can be obtained from $x$ by
trimming the last letter.
This can be expressed by the if-then-else term
\[
\treeParent(x) \defeq
\ite(x = \epsilon \lor \treeNegSpine(x),
  x \cdot 0,
  \trim_1(x)
),
\]
where the $\trim_1(x)$ function can be defined
by the formula
$
\forall x,y. y = \trim_1(x) \liff
\parens*{
  y = x \cdot 0 \lor \dots \lor y = x \cdot \ell
}$.

For $\treeChild_j(x)$ where $j \geq 2$
we simply append the letter $j$ to $x$,
and for $\treeChild_1(x)$ we need again to consider
whether or not we are on the negative spine.
If we are, the first child is obtained by trimming the last letter,
otherwise, by appending $1$:
\[
\treeChild_j(x) \defeq \begin{cases}
  \ite(\treeNegSpine(x), \trim_1(x), w \cdot 1)
  & \tif j=1,\\
  x \cdot j &\tother.
\end{cases}
\]
We can see that there is
a nice symmetry between $\treeParent$
and $\treeChild_1$,
which reflects the fact that together they form
the entire (positive and negative) spine of the tree.

Finally,
$\treeSibling_j(x)$ requires additional considerations
to ensure universal finite embeddability
and the observance of mixed order atoms.
Consider a case where we have atoms
$f(x) \llt x, f(x) \llt g(x)$
but crucially no $x \llt g(x)$ or $g(x) \llt x$:
we need to ensure that $x$ and $g(x)$
will be incomparable, regardless of
where $x$ falls in the tree.
For an explicit element
where $x$ represents a first child,
we can take $g(x)$ to be a second child
of the same parent,
which will always result in an incomparable vertex.
For explicit elements where $x$ represents a second child,
we swap the roles
and have $g(x)$ be a first child.
In general, for siblings,
we can compose ``$\treeParent$''
with an if-then-else term that
replaces $\treeChild_j$ with 
$\treeChild_1$
according to whether or not the vertex in question
is a $j$\textsuperscript{th} child.
For $j=1$ this can be done by checking
if the vertex is on the
(positive or negative) spine.
For other values of $j$,
we check whether the last letter
is $j$:
\[
\treeIsChild_j(x) \defeq \begin{cases}
  \nregexp{(0|1)\R*}(x)
  & \tif j = 1, \\
  x = \trim_1(x) \cdot j & \tother.
\end{cases}
\]

Finally, we define the terms $\treeSibling_j$ as follows:
\[
  \treeSibling_j(x) \defeq
  \ite(\treeIsChild_j(x),
    \treeChild_1(\treeParent(x)),
    \treeChild_j(\treeParent(x))
  ).
\]

Given the above terms,
for every node $n$ and function group $G$,
any path from the anchor
to some non-ground term $f(x)$ in $G$
can be composed to produce some $p^f_n(x)$,
where the free variable $x$ denotes the anchor.
For the group where the anchor is $x$,
we use the term $t^f_n \defeq p^f_n(x_1)$
($x_1$ representing an
explicit element of the first argument of $f$,
i.e., the explicit element of $n$),
otherwise $t^f_n \defeq p^f_n(\epsilon)$.

\begin{lemma}
  \label{thm:pref-well-defined}
  The constructed symbolic structure
  is well-defined.
\end{lemma}
\Proof{thm:pref-well-defined}{
  Similarly to \Cref{thm:tot-well-defined}
  for regular nodes.
  For functions
  with target summary nodes that are in linear segments,
  the mixed order atoms must always induce a linear order,
  which means all finite paths in the universal tree
  will only use the letter `0' and `1'
  and produce words within the regular language
  $\nregexp{1\R*|0\R+}$.
  For target spanning nodes,
  the terms defined in \Cref{sec:dec-pref-terms}
  will always produce words
  that represent valid locations in the universal tree.
}

\begin{lemma}
  \label{thm:pref-mixed-atoms}
  All mixed order atoms are observed.
\end{lemma}
\Proof{thm:pref-mixed-atoms}{
  Follows by case analysis.
}

This completes the proof
and the construction of $S$, thus:

\begin{corollary}
  $\explic(S) \models \pref \land \varphi$.
\end{corollary}

\begin{remark}[Complexity analysis of {$\osc*[\pref]$}]
  As previously explained,
  the construction of a symbolic model
  for $\osc*[\pref]$
  provides a bound on the size of the model 
  and identifies finite sets of bound formulas, 
  function terms and relation formulas,
which
  may be computed from the formula.
Note however that the decision procedure
  induced by the symbolic model property
  reduces satisfiability of formulas
  in $\osc*[\pref]$
  to validity of formulas in $\str$,
  which is non-elementary.
  Thus the exact size bounds of symbolic structures
  are non-important for the complexity analysis.
\end{remark}

\todo{expressivity of prefix order vs total order}
 
\subsection{Decidability of 
  \texorpdfstring{$\osc*[\total\land\psucc]$}{OSC*[TOT+PROSUCC]}
  and 
  \texorpdfstring{$\osc*[\total\land\rpred]$}{OSC*[TOT+REGPRED]}
}
\label{sec:dec-total-variants}
One notable property of the symbolic structures
constructed in \Cref{sec:dec-total,sec:dec-pref}
is that some elements have neither successors nor predecessors.
In particular, both in the case of a total order
and of a prefix order, 
regular nodes --- i.e., interpretations of constants ---
may be ``disconnected'':
since summary nodes stretch from negative infinity 
to positive infinity,
regular nodes adjacent to summary nodes
have no immediate neighbors.

In this section and the next we consider
two variants of total and prefix orders
where the order relation $\llt$ is axiomatized
such that there must exist a successor 
(in one variant)
or a predecessor (in the other).
To make these orders viable,
we also restrict all functions to 
progressive or respectively regressive,
so as to ensure mixed order atoms
would not force an infinitely 
decreasing (respectively increasing)
summary nodes.
The induced internal shapes of summary nodes
are then unidirectionally infinite,
which ensures the existence 
of successors or predecessors.

The adapted construction for 
$\osc*[\total\land\psucc]$
(respectively $\osc*[\total\land\rpred]$)
reuses most of the construction of \Cref{sec:dec-total}.
However, one key adjustment is needed.
Namely,
instead of having $\top$ as the bound formulas
for summary nodes,
we define them to
$\bound(n) = x \geq 0$ in the $\total\land\psucc$ variant,
and $\bound(n) = x \leq 0$ 
in the $\total\land\rpred$ variant.
This matches the desired internal shapes,
and ensures that all elements 
have immediate neighbors,
thus satisfying the successor/predecessor axioms.
Importantly, well-definedness is preserved:
in $\osc*[\total\land\psucc]$,
progressivity ensures 
that for any node $n$, the mixed order atoms
will induce an order where the anchor $x$ is minimal
among all non-ground terms,
and as a result, all function terms 
$t^f_n$ will be of the form $x_1 + k^f_n$
for some non-negative integer $k^f_n$.
Similarly in $\osc*[\total\land\rpred]$,
regressivity ensures all function terms
to be of the form $x_1 + k^f_n$
for some non-positive integer $k^f_n$.

\begin{lemma}
  \label{thm:tot-variants}
  The constructions for 
  $\osc*[\total\land\psucc]$
  and $\osc*[\total\land\rpred]$
  are well-defined
  and correct.
\end{lemma}
\Proof{thm:tot-variants}{
  We give the main argument for $\psucc$,
  the reasoning for $\rpred$ is symmetrical.
  Since all functions are progressive,
  all function terms in the symbolic structure
  will be $x + k$ for some \emph{non-negative} $k$.
  Since for every node $n$ we have
  $\bound(n) \models x >= 0$
  (including regular nodes),
  it follows that all terms will be in bounds.
}

\subsection{Decidability of  
  \texorpdfstring{$\osc*[\pref\land\psucc]$}{OSC*[PREF+PROSUCC]}
  and 
  \texorpdfstring{$\osc*[\pref\land\rpred]$}{OSC*[PREF+REGPRED]}
}
\label{sec:dec-pref-variants}

Similarly to the total order variants,
the internal shapes of summary nodes
become unidirectionally infinite
for the variants of $\osc*[\pref]$,
and we adapt the bound formulas 
and function terms 
to reflect this.

For the successor variant
$\osc*[\pref\land\psucc]$,
the bound formulas for linear summary nodes
are $\nregexp{1\R*}(x)$,
representing a positive line,
and for spanning summary nodes they are
$\nregexp{(1\R-$\ell$)\R*}(x)$,
representing a positive rooted tree.
When defining the function terms,
in order to preserve well-definedness,
we avoid $\treeParent(x)$ and $\treeSibling_j(x)$ terms.
To that end, we always include the term $x$
as the anchor for all function groups
and construct paths from it.
Due to progressivity, $x$ will always be 
the minimum among the non-ground terms,
and thus paths will only use 
$\treeChild_j(x)$ terms.

For the predecessor variant
$\osc*[\pref\land\rpred]$,
the predecessor axiom requires
all summary nodes to have the internal shape
of a negative line,
using the bound formula
$\nregexp{0\R*}(x)$ for all summary nodes.
Similarly to $\pref\land\psucc$,
when defining function terms
we always include the term $x$ as the 
anchor for all function groups,
but as a mirror-image of 
$\pref\land\psucc$,
here $x$ will always be the maximum
among the non-ground terms,
and all paths will only use
$\treeParent(x)$ terms.
In fact, in this case we can use 
the theory of \lia{} to express bound formulas
and function terms.
But notably, unlike in the symbolic structures
constructed for $\osc*[\total\land\rpred]$,
the overall order interpretation in
the symbolic structures for 
$\osc*[\pref\land\rpred]$
may still be non-linear,
due to the order between constants.

\begin{lemma}
  \label{thm:pref-variants}
  The constructions for
  $\osc*[\pref\land\psucc]$
  and $\osc*[\pref\land\rpred]$
  are well-defined and correct.
\end{lemma}
\Proof{thm:pref-variants}{
  Analogously to \Cref{thm:tot-variants},
  the progressive/regressive axioms
  ensure well-definedness,
  and observance of the mixed order atoms
  follows \Cref{thm:pref-mixed-atoms}.
}
  \section{Implementation}
\label{sec:impl}

We implemented a tool for 
exploring theory-generic
symbolic structures by building on the
open-source \fest{}~\cite{infinite-needle} 
Python library, 
which originally supported
symbolic structures over \lia{}.
We extended \fest{} to a
generic framework,
enabling model-checking of symbolic structures 
over different decidable theories. 
As a concrete instantiation beyond
\lia{}, we added the $\str$ 
theory of strings,
whose decision procedure
is implemented via a translation to Weak Monadic
Second-Order Logic of One Successor (\wsis).
The latest version of \fest{} is available at~\cite{elad26-fest}.

To support this translation, 
we developed Python bindings to
the \mona~\cite{mona} solver for \wsis{} 
and integrated them with the
generic \fest{} infrastructure. 
Implementing the decision procedure
for \str{} by translation to \wsis{}
allows us to efficiently support
common string operations
via a specialized, optimized translation.
For example, we support
$\trim_1$ and $\prefOf_0$,
as well as specific
regular expressions like $\nregexp{0\R*}$
(all of which are used in the symbolic structures
constructed in \Cref{sec:dec-pref}).

Our implementation serves
as a proof of concept,
demonstrating the feasibility 
of our approach and providing
a preliminary platform for experimenting 
with non-\lia{} symbolic structures.
 \section{Conclusion and Future Work}
\label{sec:conclusion}

We have presented a generalization of symbolic structures to
arbitrary base theories with a standard model, 
and have shown how this formalism 
can be used to establish decidability
results for fragments of \fol{}
that do not enjoy
a finite model property. 
Our main technical contribution is
a symbolic model property for several fragments in the
Ordered Self-Cycle family, 
obtained via a generic construction 
of symbolic models from arbitrary models. 
This yields new decidability results, 
including for the Prefix-Ordered Self-Cycle fragment, 
and extends prior
work that was limited to total orders 
and symbolic structures over \lia.

The framework developed in this paper opens several
directions for further inquiry. 
On the logical side, it would be interesting 
to identify additional
decidable fragments using symbolic model properties, in
particular fragments based on other kinds 
of order relations,
such as partial orders or lattice-like structures.

On the practical side,
while our focus has been on foundational results,
symbolic structures are intended as a means for
counter-model generation in verification. 
An important direction for future work is 
to enhance the toy implementation of 
symbolic structures
into a fully-fledged tool,
and evaluate its effectiveness
on realistic verification benchmarks. 
Understanding how symbolic counter-models can be
presented to users and integrated 
into existing verification workflows 
remains an open and relevant challenge.
 
\clearpage
\bibliography{references.bib}

\ifarxiv
\appendix
\crefalias{section}{appendix}
\section{Proofs}
\label{sec:proofs}
\Proofs{} \fi

\end{document}